\documentclass[sigconf, nonacm]{acmart}

\newcommand\vldbdoi{XX.XX/XXX.XX}

\newcommand\vldbavailabilityurl{https://github.com/sijielan/FDP_Benchmarking.git}
\usepackage{tabularx}

\usepackage{tikz}
\usepackage{amsmath}
\usepackage{filecontents}
\usepackage{comment}

\usepackage{graphicx}
\usepackage{subcaption}
\usepackage{ifthen}

\usepackage{xcolor}
\usepackage{microtype}
\usepackage{todonotes}
\usepackage{microtype}
\usepackage{enumitem}
\setlist[itemize]{nosep}
\setlist[enumerate]{nosep}

\newcommand{\minisec}[1]{\noindent\textbf{\textit{#1.}}}

\usepackage{threeparttable}
\newboolean{isMainPaper}
\setboolean{isMainPaper}{false} % Set to true for main paper and false for paper with appendix
\newcommand{\conditionaltext}[2]{
    \ifthenelse{\boolean{isMainPaper}}
        {#1} % Code to use if TRUE
        {#2} % Code to use if FALSE
}
\newcounter{takeawayctr}
\newcommand{\takeaway}[1]{%
  \refstepcounter{takeawayctr}%
  \vspace{0.35em}\noindent
  
  \textit{\textbf{Takeaway~\thetakeawayctr:}~#1}%
  \vspace{0.35em}%
}

\begin{document}
\title{FDP: The Data Placement Promise of Modern NVMe SSDs 
%[Experiment, Analysis \& Benchmark]
}

%%
%% The "author" command and its associated commands are used to define the authors and their affiliations.
\author{Sijie Lan}
\affiliation{%
  \institution{Pennsylvania State University}}
\email{sml6844@psu.edu}

\author{Hui Qi}
\affiliation{%
  \institution{Samsung}}
\email{hui81.qi@samsung.com}

\author{Xing He}
\affiliation{%
  \institution{Samsung}}
\email{xing1.he@samsung.com}

\author{Mahmut Kandemir}
\affiliation{%
  \institution{Pennsylvania State University}}
\email{mtk2@psu.edu}

\author{Javier Gonz{\'{a}}lez}
\affiliation{%
  \institution{Samsung}}
\email{javier.gonz@samsung.com}

\author{Vivek Shah}
\affiliation{%
  \institution{Samsung}}
\email{vi.shah@samsung.com}

\begin{abstract}
NVMe SSDs are now widely deployed as the storage tier in data centers. As SSDs have evolved over the past decade, the community has continued to debate the interfaces they expose and how operating systems and storage systems should exploit them. The NVMe Flexible Data Placement (FDP) feature is the latest point in this design space. FDP introduces an interface based on Reclaim Units that enables explicit data placement to reduce device write amplification without the software engineering costs of sequential-write constraints and host garbage collection of previous approaches (e.g., OCSSD, ZNS). FDP-enabled SSDs are available in multi-vendor commercial products and deployed at scale in several data centers. Their compatibility with conventional block I/O allows existing applications to run unchanged, allowing a frictionless adoption in industry.

This paper presents an experimental evaluation of FDP SSDs to characterize their data placement guarantees over the raw device interface. We then revisit two widely deployed and distinct open-source storage systems, MySQL and RocksDB, and examine whether lifetime-based data separation and distinct write patterns built into their architectures can be mapped onto FDP SSDs without invasive changes. Our evaluation shows end-to-end WAF reductions at higher device utilization, along with QoS and throughput improvements under synthetic and real-world workloads. These results demonstrate that FDP provides a practical and deployable cross-layer mechanism for data placement with open-source ecosystem support on Linux. They also highlight why FDP is gaining traction in industry as the de facto data placement technology.
\end{abstract}

\maketitle

% %%% VLDB block start %%%
% \pagestyle{\vldbpagestyle}
% \begingroup\small\noindent\raggedright\textbf{PVLDB Reference Format:}\\
% \vldbauthors. \vldbtitle. PVLDB, \vldbvolume(\vldbissue): \vldbpages, \vldbyear.\\
% \href{https://doi.org/\vldbdoi}{doi:\vldbdoi}
% \endgroup
% \begingroup
% \renewcommand\thefootnote{}\footnote{\noindent
% This work is licensed under the Creative Commons BY-NC-ND 4.0 International License. Visit \url{https://creativecommons.org/licenses/by-nc-nd/4.0/} to view a copy of this license. For any use beyond those covered by this license, obtain permission by emailing \href{mailto:info@vldb.org}{info@vldb.org}. Copyright is held by the owner/author(s). Publication rights licensed to the VLDB Endowment. \\
% \raggedright Proceedings of the VLDB Endowment, Vol. \vldbvolume, No. \vldbissue\ %
% ISSN 2150-8097. \\
% \href{https://doi.org/\vldbdoi}{doi:\vldbdoi} \\
% }\addtocounter{footnote}{-1}\endgroup
% %%% VLDB block end %%%

% \ifdefempty{\vldbavailabilityurl}{}{
% \vspace{.3cm}
% \begingroup\small\noindent\raggedright\textbf{PVLDB Artifact Availability:}\\
% The source code, data, and/or other artifacts have been made available at \url{\vldbavailabilityurl}.
% \endgroup

\section{Introduction}
 NVMe SSDs have become the de facto persistent storage medium in modern datacenters due to their performance-cost trade-offs compared to DRAM and hard disk drives~\cite{storage-review:ZhangBPLLSKCNCK25}. However, the conventional block interface~\cite{ssd_agarwal} obscures internal flash management, preventing applications from conveying data placement information that is critical for reducing device write amplification factor (WAF) and improving device lifetime. Over the past decade, both academia and industry have explored alternative interfaces~\cite{storage-review:ZhangBPLLSKCNCK25}, ranging from fully host-managed Open Channel SSDs~\cite{lightnvm:BjorlingGB17} to partial host-managed designs such as Zoned Namespaces (ZNS)~\cite{zns:BjorlingAHRMGA21}, which improve data placement at the cost of software complexity.

Flexible Data Placement (FDP) SSDs~\cite{fdp_tp, cachelib-fdp:AllisonG0HKN025, SongQ0BNL26:warp} represent a new point in this design space based on lessons learned from ZNS. FDP exposes placement primitives that allow hosts to group writes with similar lifetimes while retaining the conventional block interface. This interface compatibility enables incremental deployment where FDP devices can replace existing SSDs without disrupting FDP-unaware applications, while allowing FDP-aware systems to exploit data placement when beneficial. Vendors such as Samsung, Kioxia, and DapuStor are targeting FDP SSDs for enterprise datacenters to reduce WAF, increase utilization, and extend device lifetime in both single-tenant and multi-tenant settings that are increasingly appearing in datacenter deployments~\cite{samsung_fdp_blog_2024,samsung_fdp_blog_2025,kioxia_fdp_gc_waf_2025,dapustor_fdp_intro_2024,dapustor_fdp_deepdive_2026}.

Stable Linux kernel support for NVMe I/O Passthru~\cite{io-passthru:JoshiG0KRGLA24} makes FDP practical for widely deployed systems via the character device using userspace NVMe commands without waiting for filesystem support. Early work has shown that modest application changes can exploit FDP to separate hot and cold data in Flash cache architectures, significantly reducing WAF and extending SSD lifetime in multiple production workloads~\cite{cachelib-fdp:AllisonG0HKN025}. The work has already been merged into CacheLib~\cite{cachelib:BergBMGGLUCBHG20} and deployed in production. These results raise the question, \textit{"Can other system classes exploit architectural data separation to reduce device write amplification and increase SSD lifetime using FDP interfaces exposed by the Linux kernel?"}.

I/O Passthru is a natural fit for CacheLib because it interacts with the SSD over the block interface. However, most systems rely on a filesystem interface and designing a filesystem wrapper for them is invasive and creates friction for production deployment. Early experimental Linux kernel support for FDP through filesystems using lifetime hints~\cite{fdp-lwn,fdp-file-io} provides an opportunity to use FDP primitives at a lower software engineering cost. \textit{This paper explores the costs and benefits of this opportunity for two widely deployed and complementary classes of data systems: LSM-tree-based key-value stores (e.g., RocksDB~\cite{DongKJS21:rocksdb-evolution}) and B-tree-based relational database systems (e.g., MySQL~\cite{botros2021highperformance:mysql}).} These systems have architectural components with distinct data lifetimes and write patterns but the conventional storage interface prevents the underlying device from exploiting it.

LSM-tree-based key-value stores incur high device WAF under write-intensive workloads and background compaction~\cite{ONeilCGO96:the-lsm,abs-1812-07527:lsm-survey}. Although LSM-tree-based architectures separate files into levels with different lifetimes, this separation is lost at the storage interface. Moreover, LSM-tree-based systems perform flash-friendly out-of-place writes during compaction by deleting files and appending data to new files, which limits their consumption of the SSD’s overprovisioned (OP) erase blocks. Our analysis shows that files within each level exhibit similar lifetimes, with lower levels being significantly hotter and smaller than higher levels. By propagating this architectural separation through FDP placement hints across both experimental filesystem and stable I/O Passthru interfaces (filesystem bypass) in the Linux kernel, we reduce RocksDB’s device WAF to approximately 1 at 90\% device utilization, nearly eliminating compaction-induced device write amplification.

In contrast, B-tree-based relational database systems~\cite{HellersteinSH07:rdbms-architecture}, such as MySQL~\cite{botros2021highperformance:mysql}, exhibit a mix of in-place and out-of-place updates across write-ahead logs, buffer managers, and double-write buffers. Our analysis shows that this mixture, together with random and sequential write patterns, leads to high device write amplification at increased device utilization. By separating these components using filesystem-level FDP hints with minimal code changes, we reduce device write amplification to \textasciitilde1.2 at 80\% device utilization and extend SSD lifetime without building a specialized DBMS-managed translation layer with exclusive SSD ownership~\cite{lee2026writessds}. Our work shows the promise of data separation alone without the costs of garbage collection on FDP SSDs.

First-generation enterprise FDP SSDs are becoming available from vendors for data center deployments, raising questions among industry practitioners and researchers about their isolation and QoS behavior. Prior work has studied FDP through emulation~\cite{SongQ0BNL26:warp}, but it has primarily focused on reproducing existing workloads~\cite{cachelib-fdp:AllisonG0HKN025} and the impact of emulation. In our characterization, we systematically evaluate the current generation of enterprise FDP SSDs~\cite{samsung_pm9d3a_2026}, ranging from 2 TB to 8 TB through the raw device interface. We quantify the limits of their isolation and QoS behavior and identify the factors that affect them under concurrent, write-intensive workloads. We hope this characterization provides a foundation for designing system abstractions for filesystems, storage systems, and database systems, and can guide further work in this area. Our end-to-end data placement approach for widely deployed open-source systems, RocksDB and MySQL, uses this characterization to exploit application-level data separation with device behavior. This approach reduces device WAF and extends SSD lifetime in single-tenant and multi-tenant settings along with improvements in application-level metrics of throughput and latency.

In this work we make the following contributions: \\
\textbf{1. Enterprise FDP SSD Characterization}. We analyze the isolation and QoS behavior of commercially available enterprise FDP SSDs for data centers and quantify the factors that affect them under concurrent, write-intensive workloads (\S~\ref{sec:microbenchmarks}). \\
\textbf{2. RocksDB Data Placement}. We show that the LSM-level architecture and compaction process of key-value stores provide data separation that maps naturally to FDP data placement. Using the filesystem interface, this mapping reduces device WAF to near-ideal levels %minimal changes
with changes to only 13 lines of code. Bypassing the filesystem through the I/O Passthru interface further improves tail latency and application-level write amplification under intense write pressure (\S~\ref{sec:rocksdb}). \\
\textbf{3. MySQL Data Placement}. We show that RDBMS components such as logging, buffer management, and double-write buffering exhibit distinct write patterns and lifetimes that can be separated using FDP. Using the filesystem hinting interface, this separation reduces device WAF, supports higher device utilization, and improves application throughput (\S~\ref{sec:mysql}). \\
\textbf{4. End-to-End Evaluation}. We perform a comprehensive evaluation of the benefits of data separation across RocksDB, MySQL, emerging and stable Linux kernel interfaces, and FDP SSDs using multiple synthetic and real-world benchmarks in single-tenant and multi-tenant settings (\S~\ref{sec:evaluation}).

\section{Background and Motivation}
\subsection{FDP In a Nutshell}
Flexible Data Placement (TP 4146) was fully ratified by the NVMe board in 2023~\cite{fdp_tp}. It originated from a joint proposal by Google and Meta that merged their SmartFTL~\cite{smartftl} and Direct Placement Mode designs. FDP extends the multi-stream SSD proposal~\cite{kang2014multi} by providing explicit data placement guarantees on SSDs while avoiding the garbage collection responsibilities imposed by ZNS~\cite{zns:BjorlingAHRMGA21}. \textbf{The proposal is fully backward-compatible, i.e., an FDP SSD can operate as a conventional SSD and does not introduce new NVMe commands}.

An FDP SSD organizes NAND flash blocks into a collection of Reclaim Units (RUs) within a reclaim group (RG). The host is given a set of reclaim unit handles (RUHs), which act as placement tags for NVMe write commands. The size of the reclaim unit is vendor-defined and may span multiple erase blocks. Emerging FDP SSDs use superblock-sized reclaim units to increase parallelism and reduce firmware indexing overheads. The SSD retains full internal control over the mapping between RUHs and reclaim units. The read commands operate without any modification. \textit{If a write command does not specify a RUH, the device applies a default reclaim unit handle}. Standard NVMe admin commands allow the host to discover the FDP topology, including reclaim groups, reclaim unit size, and the number of available reclaim unit handles.

Unlike ZNS, logical block addresses are not statically bound to particular RUHs. Instead, they are {\em dynamically} assigned as the host issues RUH tagged NVMe write commands. While FDP removes host control over garbage collection, it provides a set of event types that capture important garbage collection-related information reported by the SSD controller. These events are recorded in an NVMe log page and can be queried by the host to evaluate placement effectiveness and integrate into feedback driven placement policies. For details on current generation FDP SSD architecture and programming model, we refer readers to whitepapers~\cite{samsung_fdp_blog_2024, samsung_fdp_blog_2025}, educational videos~\cite{allison2024fdp}, and research papers~\cite{cachelib-fdp:AllisonG0HKN025, storage-review:ZhangBPLLSKCNCK25}.

\subsection{FDP Linux Support}
\label{sec:fdp_linux_support}
\begin{figure}[t]
\centering
\includegraphics[width=0.7\columnwidth]{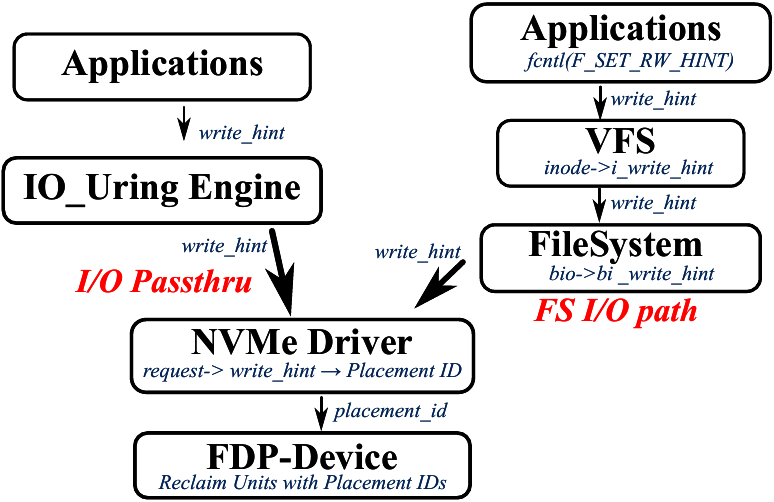}
\captionsetup{belowskip=0pt, aboveskip=2pt}
    \caption{FDP write hint propagation through I/O Passthru and filesystem paths in the Linux kernel.}
% \vspace{-3ex}
\label{fig:fdp_arch}
\end{figure}

The FDP proposal is being driven by hyperscalers, in contrast to earlier proposals that were primarily led by SSD vendors. As a result, supporting FDP within the Linux open-source ecosystem has been a priority, drawing on lessons from prior efforts, especially the Multi-Stream proposal~\cite{kang2014multi}. Applications can send data placement hints from userspace in Linux through the following mechanisms (see Figure \ref{fig:fdp_arch}).\\
\minisec{1. Using the raw character device abstraction and I/O Passthru}
An application can access an FDP SSD through \textit{a raw character device} registered by the kernel. As shown in the "I/O Passthru" path in Figure~\ref{fig:fdp_arch}, the application constructs NVMe write commands with RUH hints using the \texttt{NVME\_RW\_DTYPE\_DPLCMT} directive type and submits them to the Linux kernel via the \texttt{io\_uring\_cmd} interface~\cite{io-passthru:JoshiG0KRGLA24}, which bypasses the Virtual File System (VFS) and block layers and forwards the command directly to the Linux NVMe driver and the SSD. This interface provides applications with full control over specifying RUH hints on a per-LBA basis. I/O Passthru support has been available in the Linux kernel since version 5.19. This mechanism was used to integrate FDP support into CacheLib over the raw character device~\cite{cachelib-fdp:AllisonG0HKN025}.\\
\minisec{2. Using the filesystem and per-file lifetime hints}
To support applications that use the Linux filesystem abstraction rather than raw device access, a recent patchset has been proposed~\cite{fdp_ext4,joshi2024nvmefdp,fdp-file-io, fdp-lwn}. The patchset enables propagation of existing per-file lifetime hints in the Linux \texttt{VFS} layer to disjoint RUHs. Applications can pass lifetime hints via the \texttt{fcntl} system call using one of the available values, namely, \texttt{SHORT, MEDIUM, LONG, EXTREME, NOT\_SET, NONE}. The hint is stored in the inode, and all I/O to the corresponding file is tagged with this hint. As shown in the "FS I/O path" in Figure \ref{fig:fdp_arch}, the inode-level hint is propagated to the NVMe driver, which maps the lifetime hint to a disjoint RUH using the \texttt{NVME\_RW\_DTYPE\_DPLCMT} directive type. If the SSD supports fewer RUHs than the number of hints, the default RUH is used. Since this mechanism operates at the VFS and NVMe driver layers, it becomes automatically available across all Linux filesystems. Although this approach is highly convenient, it limits the granularity of data placement to the file level rather than individual LBAs. The mechanism is intended to allow applications to adopt FDP SSDs seamlessly through filesystem interfaces while still enabling deathtime-based grouping. 

\subsection{The Flexible Data Placement Opportunity}
In this paper, we evaluate the potential of architectural data separation through FDP using two widely deployed storage systems with distinct architectures: \\
1. \textbf{LSM tree based key-value stores} (\S~\ref{sec:rocksdb-background}) \\
2. \textbf{B-tree based relational database systems} (\S~\ref{sec:mysql-background})\\

\subsubsection{Key-Value Stores using LSM tree-based architectures (RocksDB)}
\label{sec:rocksdb-background}
\begin{figure}[htbp]
    \centering
    \includegraphics[width=0.9\linewidth]{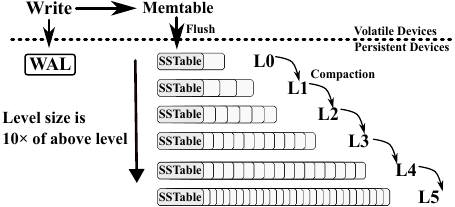}
    \caption{RocksDB Architecture Overview.}
    \label{fig:rocksdb_arch}
  \vspace{-2ex}
\end{figure}

Key-value stores such as \textsc{RocksDB}~\cite{DongKJS21:rocksdb-evolution} use log-structured merge trees (LSM trees)~\cite{abs-1812-07527:lsm-survey,ONeilCGO96:the-lsm}. As shown in Figure~\ref{fig:rocksdb_arch}, each write is appended to a \textit{write-ahead log (WAL)} and inserted into an in-memory \textit{MemTable}. When the MemTable fills, RocksDB flushes it as an immutable sorted string table (SST) file to level $L0$. Background \textit{compaction} merges SST files and propagates their contents to deeper levels ($L1 \dots Ln$), discarding obsolete entries when possible. WAL and SST files are written sequentially and obsolete files are later deleted. Although this out-of-place write pattern is conceptually SSD friendly, the lack of data separation at the interface results in data mixing at the SSD level between the different LSM levels and the WAL. Moreover, under leveled compaction the level size increases geometrically (size multiplier) across levels. Lower-numbered levels are therefore smaller and experience higher file deletion and creation rates than deeper levels.

Moreover, in leveled compaction strategies, the sizes of levels vary exponentially. Consequently, level L0 is significantly smaller in size and more frequently updated than level L1. This property holds throughout the level structures.

\minisec{SST Lifetime Analysis}
To test whether SST files can be grouped by expected deletion (death) time, we measured file lifetimes across LSM-tree levels. We instrumented RocksDB to record the creation and deletion timestamps of each SST file during workload execution. The interval between these events gives the observed lifetime of each deleted file. Across multiple five-hour runs of synthetic, compaction-intensive write-heavy workloads, files in $L0$–$L3$ had short, relatively uniform lifetimes. Most persisted for seconds to minutes before being selected for compaction. File lifetimes diverged from $L4$ onward. Files in $L4$ persisted for hundreds to tens of thousands of seconds, while many files in $L5$ remained present at the end of each run. These results indicate a marked placement boundary at $L4$, suggesting that files in $L0$–$L3$ and files in $L4$–$L5$ should be assigned to separate placement groups. This separation boundary is a function of the architecture and not the specific workload. 

\emph{In this work, we study whether separating the WAL and LSM-tree levels through the filesystem and I/O Passthru interfaces is sufficient to reduce device WAF at high utilization on FDP SSDs.}

%with out of place appends to SST files organized into different levels that are read and merged by a background compaction process~\cite{abs-1812-07527:lsm-survey, ONeilCGO96:the-lsm}. Due to the append only nature of these leveled SST files, followed by merging and deletes during compaction, the resulting I/O patterns consist primarily of sequential writes and LBA deallocations that are friendly to SSDs and allow the device to minimize internal data movement.  In this work, we evaluate whether data mixing across tiered LSM levels is problematic and how it can be alleviated using FDP non-invasively. 
%\noindent RocksDB uses a Log-Structured Merge-tree (LSM-tree) architecture. 

\subsubsection{B-Tree based Relational Database Systems (MySQL)}
\label{sec:mysql-background}
\begin{figure}[t]
  \centering
  \includegraphics[width=0.8\columnwidth]{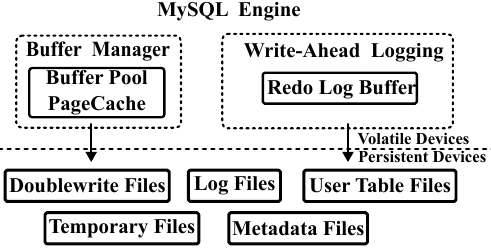}
  \caption{Data classes written by the MySQL/InnoDB engine. All writes are
  staged through the in-memory buffer pool and then land on storage as user
  tables, metadata, the double-write buffer, redo logs, and temporary files.}
  \label{fig:mysql-arch}
  %\vspace{-2ex}
\end{figure}

Relational database systems such as \textsc{MySQL}~\cite{botros2021highperformance:mysql} rely on B-tree (and its variants) to organize table data and indexes into pages managed by a buffer pool~\cite{HellersteinSH07:rdbms-architecture}. While updates modify pages in memory, dirty pages are later flushed to their locations in-place in the data files, resulting in random writes on SSDs. To support crash recovery, MySQL's InnoDB storage engine records changes in a sequentially written redo log analogous to a write-ahead log in a key-value store. InnoDB writes pages being flushed to a doublewrite buffer before writing them to their final locations in the data files. This protects against torn page writes during a crash.

Percona Server for MySQL is an open-source, performance-enhanced drop-in replacement for MySQL Community Edition~\cite{PerconaServer2023} that is widely deployed in high-throughput transactional environments. As shown in Figure~\ref{fig:mysql-arch}, it uses InnoDB as its default storage engine, whose storage files fall into five categories: (1) user table files, (2) log files, (3) temporary table files, (4) doublewrite buffer files, and (5) metadata files~\cite{MySQLInnoDBArch2024}. These categories exhibit distinct lifetime and write-pattern characteristics, making them natural candidates for differentiated data placement. User table files are persistent and receive predominantly random in-place updates, whereas redo log files are sequentially written and cyclically reused. Temporary table files are short-lived, while doublewrite buffer files are fixed in size and cyclically overwritten at high frequency. Metadata files are generally smaller and exhibit different update characteristics from user data. Our empirical characterization reveals that user table and doublewrite buffer files dominate the write traffic, accounting for 44.9\% and 43.7\% of host writes, respectively. In contrast, metadata and log files account for 7.0\% and 4.3\%, respectively, while temporary files contribute less than 0.1\%.

%\todo[inline] {Add the table text of writes for MySQL}

Concurrent updates from these storage components lead to mixing of data with different lifetimes in the same SSD erase blocks causing increased device-level write amplification. \emph{In this work, we study whether MySQL's existing separation of storage files can be used to assign distinct lifetime hints through the filesystem interface, reducing device-level write amplification on FDP SSDs at high utilization.} The required data placement changes comprise only 31 lines of code across the source files.

\subsubsection{\textbf{Why is SSD write amplification important \emph{now}?}}
The lifetime or endurance of a NAND Flash-based SSD is largely determined by its \emph{device write amplification factor (WAF)} due to the limited number of program/erase cycles and log-structured overwrite nature of Flash cells~\cite{storage-review:ZhangBPLLSKCNCK25, ssd_agarwal, wu2020bitflip}.  Enterprise SSD deployments in datacenters are actively provisioned and monitored so that the device WAF remains close to \textasciitilde1. Even modest increases in WAF can significantly reduce device lifetime and quality of service (QoS), directly translating into higher infrastructure costs at scale. Recent SSD supply constraints and rising NAND Flash costs have forced hyperscalers to maximize usable capacity, making capacity overprovisioning an unattractive mechanism for controlling WAF~\cite{reutersaifrenzy, trendforce2025enterprisessd}. Moreover, lowering WAF also improves datacenter sustainability~\cite{embodiedcarbon, microsoft-sustainability, google-sustainability, meta-sustainability, amazon-sustainability} and it also correlates with \emph{performance stability and SSD power consumption}. Excessive garbage collection not only wears out flash cells and consumes power, but also increases latency variability and contributes to tail latency spikes which are particularly harmful for latency-sensitive services~\cite{LiPSKLDKGG23:io-determinism}. 

\vspace{-1ex}
\section{The FDP Promise of the Raw Device}
\label{sec:microbenchmarks}
In this section, we evaluate the isolation behavior of FDP SSDs and their impact on device WAF, I/O throughput (IOPS), and latency under heavy concurrent write patterns that lead to high device WAF at full device utilization. This characterization establishes the potential benefits and limitations of separating data by lifetime at the application layer.\\  \textbf{Tools and Setup.} We use FIO (v3.35)~\cite{fio} with the \texttt{io\_uring\_cmd} NVMe engine via \texttt{libnvme} v1.16.1 on Linux kernel 6.9, which supports I/O Passthru to generate synthetic write workloads. Table~\ref{tab:fio-all} summarizes the microbenchmark configurations. We use the character device in these experiments to minimize any interference effects of the Linux kernel and its I/O stack. Our experiments are conducted on three PM9D3a FDP SSDs~\cite{samsung_pm9d3a_2026} with capacities of 1.88 TB, 3.84 TB, and 7.68 TB, each exposing 8 RUHs with reclaim unit sizes of approximately 6.44 GB, 13.08 GB and 17.54 GB, respectively. Due to space constraints, we present results for the 3.84 TB device in the main paper and provide a comparison across device capacities in\conditionaltext{ the extended paper in supplementary materials.}{\S~\ref{sec:app-fdp-device-sizes} (supplementary material)}.  We obtain device WAF using vendor-specific logs for accuracy\footnote{WAF can also be measured using the NVMe OCP commands.}. \emph{All workloads are evaluated on the same FDP device in two modes, (1) with FDP enabled (\textbf{FDP}), and (2) with FDP disabled for the conventional NVMe SSD baseline (\textbf{Conv.})}. In the FDP mode, workloads are assigned to a RUH while in the Conv mode they are not.

Our experiments address questions raised by datacenter operators about FDP SSD behavior under common workloads and deployment conditions:\\
\emph{1. How effective is RUH isolation under concurrent sequential workloads (\S~\ref{sec:seq-workloads})?\\
2. How effective is RUH isolation when RUHs are shared between concurrent writers (\S~\ref{sec:shareruh})?\\
3. How effective is RUH isolation under a mix of concurrent sequential and random workloads (\S~\ref{sec:randvsseq})?\\
4. How effective is RUH isolation in the presence of mixed hot and cold data (\S~\ref{sec:microbench:hot-cold})?\\
5. How effective is RUH isolation when concurrent workloads have mixed skewed and uniform access patterns (\S~\ref{sec:workload-skew})?}\\

\begin{table*}[t]
\centering
\small
\caption{FIO microbenchmark configurations. }
% All experiments use the \texttt{io\_uring\_cmd} NVMe passthrough engine on a 3.84\,TB NVMe SSD; in FDP mode each job (or group of jobs sharing a region) is mapped to its own RUH.}
\label{tab:fio-all}
\begin{tabularx}{\textwidth}{@{}l c >{\raggedright\arraybackslash}X c l@{}}
\toprule
\textbf{Experiment} & \textbf{Jobs} & \textbf{Access pattern} & \textbf{Block size} & \textbf{Run length / stop condition} \\
\midrule
\S3.1 Sequential (uneven / even) & 8 & Sequential write; ranges 2--23\,\% / 12\,\% each & 128\,KiB & 8\,h (time-based) \\
\S3.2 Shared RUHs & 8--64 & Sequential write; 1, 2, 4, 8 jobs per RUH & 128\,KiB & 8\,h (time-based) \\
\S3.3 Varying randomness & 8 & 8S $\rightarrow$ 8R; smallest ranges become random & 128\,KiB & 8\,h (time-based) \\
\S3.4 Hot/cold prefill & 1 & Random write after 50--99\,\% cold prefill & 4\,KiB & $10\times$ device capacity written \\
\S3.5 Skew, single & 1 & Zipfian random write (iolog), $z = 0$--$1.4$ & 16\,KiB & $10\times$ device capacity written \\
\S3.5 Skew, co-running & 2 & Uniform + Zipfian random write, 3 LBA splits & 4\,KiB & 8\,h (time-based) \\
\bottomrule
\end{tabularx}
\end{table*}

\subsection{RUH Isolation With Sequential Workloads}
\label{sec:seq-workloads}
\textbf{Setup.} To quantify the best-case isolation benefits of data placement on FDP, we run eight FIO sequential write workloads over disjoint LBA ranges, modeling a perfectly partitioned workload stressing the SSD while maintaining a constant invalidation rate of LBAs. In the FDP mode, each workload is pinned to a distinct RUH. Each workload writes sequentially within its assigned range and wraps around upon reaching the end. Each experiment runs for 8 hours to evaluate long-term behavior and resilience. We partition the device using two policies:\\
(1) \textbf{Even Partitioning.} The device is divided into eight equal-sized, non-overlapping regions, each assigned to one write stream. This setup isolates interference under balanced spatial partitioning, similar to ZNS-friendly workloads~\cite{zns:BjorlingAHRMGA21}.\\
(2) \textbf{Uneven Partitioning.} To model heterogeneous tenant allocations, we assign progressively larger regions to each workload, \emph{ranging from 2\% to 23\% of device capacity in 3\% increments, summing to 100\%}. This creates non-uniform utilization and stresses the SSD’s ability to preserve isolation under heterogeneous spatial utilization.\\

\begin{figure}[t]
    \centering
    \includegraphics[width=\columnwidth]{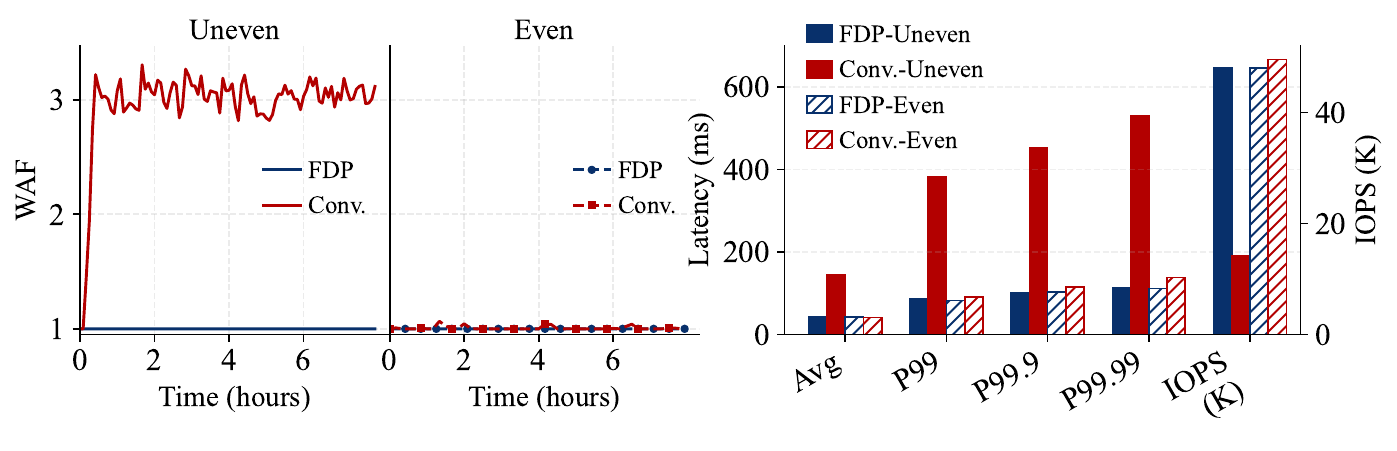}
    \captionsetup{belowskip=0pt, aboveskip=2pt}
    \caption{FDP vs.\ Conv. SSD WAF, latency, and IOPS under eight sequential streams with even and uneven partitioning.}
 % \vspace{-2ex}
    \label{fig:compare_sequential}
\end{figure}

\textbf{Results.} Figure~\ref{fig:compare_sequential} shows WAF, latency, and throughput. Under even partitioning, both FDP and Conv exhibit an ideal WAF of \textasciitilde1 and similar latency and IOPS. Since logical regions are uniformly sized and distributed, writes spread uniformly across erase blocks at similar rates, resulting in minimal internal SSD data movement. In contrast, clear divergence appears under uneven partitioning. While Conv WAF increases to \textasciitilde3, FDP maintains a WAF of \textasciitilde1. Here, invalidation rates (due to wrap-around overwrites) diverge across concurrent write streams, leading to uneven invalidation within erase blocks in Conv. With FDP RUH hinting, streams with similar invalidation rates are isolated, preserving behavior similar to even partitioning. In this setting, FDP reduces average latency by 70\% and tail latencies (P99–P99.99) by approximately 77\% by mitigating GC interference, while improving IOPS by 3.3$\times$. \textit{Uneven partitioning is common in production systems, where multi-tenant workloads exhibit heterogeneous data footprints and growth rates.}
\takeaway{
FDP provides strong isolation of concurrent sequential writers under both even and uneven partitioning, maintaining near-ideal WAF and reducing average and tail latencies by ~70\% and 77\%, whereas conventional SSDs suffer from GC-induced interference.
}

\subsection{RUH Isolation With Shared Sequential Workloads}
\label{sec:shareruh}
\textbf{Setup.} To stress the benefits of FDP isolation when a RUH is shared by multiple sequential write workloads, we increase the number of workloads mapped to each RUH under \emph{uneven partitioning} by subdividing each per-RUH workload into multiple non-overlapping sub-workloads and running them concurrently. Specifically, the uneven workload assigned to each RUH in \S~\ref{sec:seq-workloads} is evenly partitioned into distinct LBA ranges while preserving the same aggregate size and access patterns. For \textbf{16 workloads}, two workloads are assigned to each RUH. This design introduces intra-RUH contention and models realistic deployments where strict one-to-one mapping between workloads and RUHs is impractical, particularly given that current devices expose a limited number of RUHs (e.g., 8). Here, we vary the total number of concurrent workloads from 8 to 64, corresponding to RUH-to-workload mappings from 1:1 to 1:8.

\begin{figure}[t]
%\vspace{-2ex}
  \centering
  \includegraphics[width=\columnwidth]{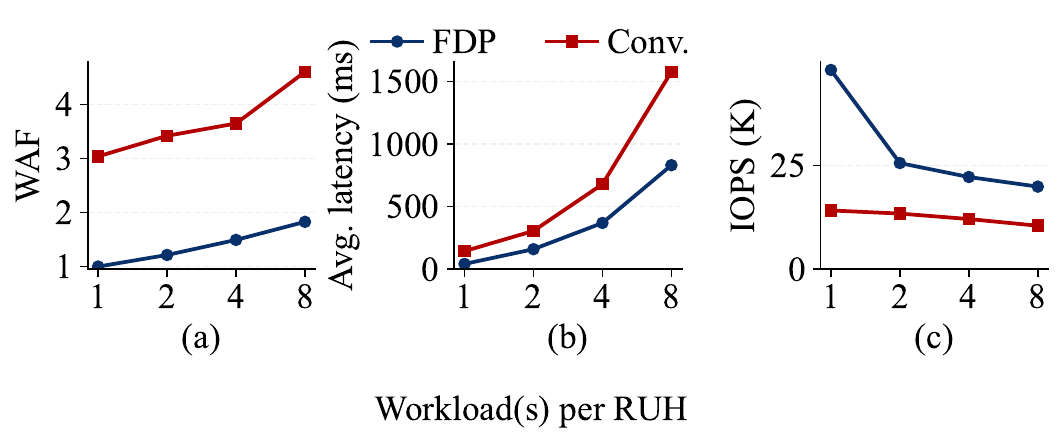} 
  \captionsetup{belowskip=0pt, aboveskip=2pt}  
  \caption{FDP vs.\ Conv SSD WAF, write latency, and IOPS with sharing of RUHs.}
  \label{fig:shareruh_scale}
\end{figure}

\textbf{Results.}
Figure~\ref{fig:shareruh_scale}(a) shows device WAF as the RUH-to-workload mapping changes from 1:1 (8 workloads) to 1:8 (64 workloads). As concurrency increases, Conv exhibits steadily rising WAF due to mismatched cross-workload invalidations within erase blocks, reaching 4.60 at 64 workloads. Under FDP, sharing an RUH mixes writes from different writers with mismatched invalidation patterns, also increasing WAF. Nevertheless, FDP's WAF remains substantially lower across all concurrency levels, peaking at 1.83 at 64 workloads. This indicates that RUH-based isolation continues to mitigate cross-workload interference even with increased sharing, and suggests that the limited number of RUHs in current FDP SSDs may not be a practical barrier to effective data separation.

Figure~\ref{fig:shareruh_scale}(b) shows the average write latency, which mirrors the throughput results. FDP lowers the average write latency relative to Conv at every sharing ratio, by about 70\% at the 1:1 mapping (42~ms vs.\ 145~ms) and by roughly 47\% even at the highest 1:8 sharing ratio (831~ms vs.\ 1577~ms). Figure~\ref{fig:shareruh_scale}(c) shows the corresponding write IOPS. FDP sustains a consistent throughput advantage over Conv across all concurrency levels, retaining a 1.9$\times$ improvement even at the highest RUH sharing ratio (19.9~K vs.\ 10.5~K IOPS at 64 workloads). This demonstrates that RUH-based isolation remains effective in reducing garbage collection interference under increased RUH sharing. We further study the impact of more sophisticated partitioning policies under shared-RUH workloads\conditionaltext{which is available in the expanded paper in the supplementary materials.}{in \S~\ref{app:homo_heter_map} (supplementary material).}

% \begin{figure}[t]
%   \begin{subfigure}[t]{0.49\columnwidth}
%     \centering
%     \captionsetup{belowskip=0pt, aboveskip=2pt}
%     \includegraphics[width=\linewidth]{figures/fio_workloads/shareRUH/waf_across_workloads.pdf}
%     \caption{WAF}
%     \label{fig:shareruh_waf_scale}
%   \end{subfigure}
%   \hfill
%   \begin{subfigure}[t]{0.49\columnwidth}
%     \centering
%     \captionsetup{belowskip=0pt, aboveskip=2pt}
%     \includegraphics[width=\linewidth]{figures/fio_workloads/shareRUH/latency_across_workloads.pdf}
%     \caption{Write Latency}
%     \label{fig:shareruh_iops_scale}
%   \end{subfigure}
%   \captionsetup{belowskip=0pt, aboveskip=2pt}
%   \caption{FDP vs.\ Conv WAF and write latency performance with sharing of RUHs.}
%   %\vspace{-4ex}
%   \label{fig:shareruh_scale}
% \end{figure}

\takeaway{FDP sustains lower WAF, higher IOPS, and lower write latency even under increased RUH sharing, indicating that effective isolation and invalidation of the OP space persist even with limited RUHs among concurrent sequential writers.}

\subsection{RUH Isolation With Mixed Random and Sequential Workloads}
\label{sec:randvsseq}
\textbf{Setup.} To stress SSD garbage collection and pressure on internal SSD OP space, we study the impact of random writes on FDP isolation under the (\emph{uneven partitioning}) setup. We run eight FIO write workloads mapped to uneven logical address ranges as in \S~\ref{sec:seq-workloads}. As before, in FDP mode, each workload is pinned to a distinct RUH. We begin with eight fully sequential workloads (\textbf{8S}) and gradually introduce random writes in the workloads within its assigned range. We incrementally convert the remaining sequential stream with the smallest assigned logical size to random writes, while keeping the other streams unchanged. This process continues until all eight streams are random (\textbf{8R}). The intermediate configuration (\textbf{4S4R}) corresponds to the four smallest regions issuing random writes and the four largest remaining sequential. \emph{This setup models realistic multi-tenant workloads, where smaller or short-lived tenants often exhibit bursty random access patterns, while larger tenants retain more sequential behavior~\cite{cachelib-fdp:AllisonG0HKN025, SongQ0BNL26:warp}}. 

% \begin{figure}[t]
%     \centering
%     \includegraphics[width=0.6\columnwidth]{figures/fio_workloads/varyrandom/waf_vary_rand.pdf}
%     \captionsetup{belowskip=0pt, aboveskip=2pt}
%     \caption{FDP vs.\ Conv WAF with uneven partitioning with a mix of random and sequential workloads.}
%     \label{fig:varyrand}
%  %\vspace{-4ex}
% \end{figure}

\begin{figure}[t]
  \centering
  \begin{minipage}[t]{0.45\columnwidth}
    \centering
    \includegraphics[width=\linewidth]{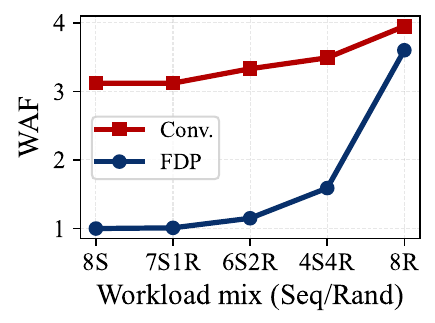}
    \captionsetup{belowskip=0pt, aboveskip=2pt}
    \captionof{figure}{WAF vs.  mix.}
    \label{fig:varyrand}
  \end{minipage}\hfill
  \begin{minipage}[t]{0.45\columnwidth}
    \centering
    \includegraphics[width=\linewidth]{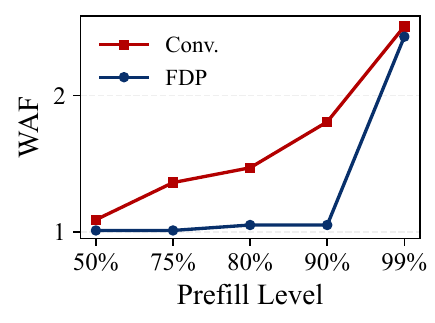}
    \captionsetup{belowskip=0pt, aboveskip=2pt}
    \captionof{figure}{WAF vs.\ prefill ratio.}
    \label{fig:prefill}
  \end{minipage}
 % \vspace{-2ex}
\end{figure}

\textbf{Results.} Figure~\ref{fig:varyrand} shows the results on device WAF. As in \S~\ref{sec:seq-workloads}, FDP achieves near-ideal WAF (\textasciitilde1) for fully sequential workloads (\textbf{8S}), whereas Conv incurs a WAF of \textasciitilde3. As randomness is progressively introduced in smaller logical regions, WAF increases for both modes. At \textbf{4S4R}, where cumulatively 26\% of the LBA range is written randomly, FDP maintains a substantially lower WAF (\textasciitilde1.59) compared to Conv (\textasciitilde3.49), a \textbf{54\%} reduction. The effectiveness of FDP decreases as random writes introduce random invalidations across erase blocks, and the overprovisioned (OP) space cannot absorb incoming writes without triggering garbage collection and internal data movement. When all streams become random (\textbf{8R}), spatial locality largely disappears, and both modes exhibit elevated WAF (3.60 vs.\ 3.95), reducing FDP’s advantage to about 9\%. These results show that FDP remains effective at isolating data streams as long as pressure on the shared OP space is limited. Under heavy GC pressure from random write workloads, shared OP space can lead to noisy neighbor interference~\cite{SongQ0BNL26:warp}.

\takeaway{
FDP exhibits perfect isolation and reduces WAF under concurrent sequential and random writes as long as the invalidations of the random writes can mostly be absorbed by the internal SSD OP space.}

%\vspace{-1ex}
\subsection{RUH Isolation With Mixed Hot and Cold Data}
\label{sec:microbench:hot-cold}
\textbf{Setup.} In many storage systems, hot (frequently updated) and cold (rarely updated) data coexist in SSDs due to application-driven access patterns~\cite{kang2014multi, SongQ0BNL26:warp, cachelib-fdp:AllisonG0HKN025, HuangBCNSSQ17:flashblox}. This effect is amplified in cloud environments running multi-tenant applications, where VMs exhibit heterogeneous activity levels. To evaluate FDP’s ability to \emph{separate} hot and cold data under challenging random write workloads, we run a single random write workload while varying the fraction of the SSD LBA range it accesses. The active region models hot data, while the remaining capacity represents cold data. We precondition the SSD by writing it to full capacity multiple times before each experiment. The \emph{prefill ratio} denotes the fraction of the total SSD capacity written during preconditioning to simulate cold data. The random-write workload then accesses the remaining $(1-p)$ portion of the LBA space, representing the hot region. A higher prefill ratio leaves less space for hot writes, intensifying the GC pressure on the active region. In the FDP mode, the active workload is mapped to a dedicated RUH, isolating hot writes from the cold prefilled region. 

% \begin{figure}[t]
%     \centering
%     \includegraphics[width=0.5\columnwidth]{figures/fio_workloads/prefill/waf_compare_line.pdf}
%     \captionsetup{belowskip=0pt, aboveskip=2pt}
%     \caption{FDP vs.\ Conv WAF under random write workloads with varying prefill ratios.}
%     \label{fig:prefill}
%  %\vspace{-3ex}
% \end{figure}

\textbf{Results.} Figure~\ref{fig:prefill} shows WAF with varying prefill ratios. With FDP, WAF remains nearly constant (around 1.0) up to 90\% prefill and increases only to 2.43 at 99\% utilization. Without FDP, WAF increases steadily with utilization, from 1.09 (50\%) to 1.81 (90\%), reaching 2.51 at 99\%. At high utilization, FDP reduces WAF by up to \textbf{44\%}. This improved resilience is due to FDP’s ability to colocate hot LBAs within erase blocks, enabling more efficient invalidation. Together with the utilization of device OP space, this allows random writes to be absorbed more effectively.

\takeaway{FDP enables higher device utilization with low WAF under hot and cold data layouts by allowing the SSD internal OP space to be effectively used by hot data.}

\vspace{-1ex}
\subsection{RUH Isolation Under Skewed Accesses}
\label{sec:workload-skew}
\textbf{Setup.} To evaluate how data locality affects device WAF and whether FDP mitigates interference under mixed workloads with different skew levels, we run experiments with two FIO write workloads, one accessing the LBA range uniformly at random and the other following a Zipfian distribution. We vary the Zipfian skew ($skew \in {1.0, 1.2, 1.4}$) and the LBA allocation between the workloads using splits of 20/80, 50/50, and 80/20.  Real world workloads often exhibit a mix of skewed and uniform access patterns due to heterogeneous application behavior, making it important to evaluate such combinations~\cite{YangZ16:skewed-workloads}.

\begin{figure}[ht]
    \centering
    \captionsetup{belowskip=0pt, aboveskip=2pt}
    \includegraphics[width=\linewidth]{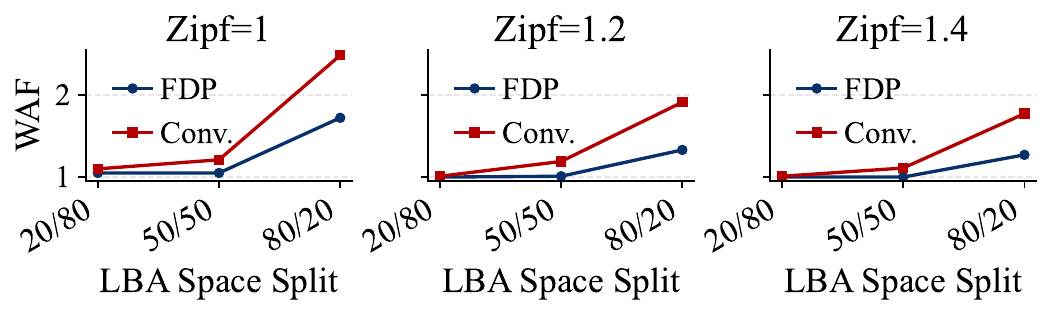}
  %  \caption{FDP vs.\ Conv WAF under mixed workloads with varying skew values and LBA space allocations.}
    \caption{FDP vs.\ Conv. SSD WAF under mixed workloads with varying Zipf skew values and LBA space splits (uniform/Zipfian).}
    \label{fig:skew_comparison}
 %\vspace{-2ex}
\end{figure}

\textbf{Results.} To set up a baseline for Zipfian access patterns, we ran a single workload with varying values of the Zipfian constant. Consistent with previous SSD WAF results with varying skew~\cite{Desnoyers14:analytic-ssd-models,HaasLBL25:ssd-iq}, our results show a strong inverse relationship between skew and WAF due to repeated accesses to LBAs resulting in lower pressure on OP space. With uniform random writes ($skew=0$), WAF rises rapidly and saturates near 4.0. In contrast, highly skewed workloads (e.g., $skew=1.4$) sustain WAF close to 1.0. Full characterization of this baseline is provided in
\conditionaltext{the extended paper in supplementary materials.}{\S~\ref{app:skew-vs-waf}(supplementary material.)}

Figure~\ref{fig:skew_comparison} shows the impact on the mixed write workload (1 random and 1 Zipfian) setup across different splits and skew values. In both Conv and FDP modes, device WAF increases as the fraction of LBA space allocated to the random workload grows, due to higher pressure on OP space from widely distributed invalidations across the LBA range. However, FDP is more resilient as it confines random and Zipfian workloads to separate LBA ranges, resulting in higher invalidation locality and lower pressure on OP space. As skew increases, FDP’s benefit becomes more pronounced. For example, at $skew=1.4$ with the \textbf{80/20} split, FDP maintains a WAF of 1.25, while Conv degrades to 1.77.

\takeaway{FDP improves WAF under mixed workloads with skews by preserving locality, with gains increasing as skew grows and random access pressure on the LBA range rises.}

\minisec{More Takeaways} \emph{Our microbenchmarks also show scalable WAF behavior across all three FDP device sizes and demonstrate effective WAF isolation for mixed workloads using filesystem temperature hints across XFS, EXT4, Btrfs, and F2FS in\conditionaltext{ the extended paper in supplementary materials.}{\S~\ref{sec:app:fs-microbench} (supplementary material).} We omit these results from the main paper due to space constraints.}

\subsection{FDP Data Placement Guidelines}

\textbf{1. Group writes by invalidation rate.}
Placement groups should contain data that becomes invalid at similar rates due to spatial and temporal locality. Under uneven sequential workloads, separating streams by RUH reduces WAF from approximately 3 to 1 because each RUH contains writes with a similar invalidation rate. 

\noindent\textbf{2. Share RUHs only among compatible write streams.}
A one-to-one mapping between writers and RUHs is not necessary. FDP maintains lower WAF and higher throughput when up to eight sequential writers share each RUH. However, WAF increases from approximately 1 to 1.83 as the sharing ratio increases from 1:1 to 1:8. RUHs should therefore be shared among streams with similar access and invalidation patterns.

\noindent\textbf{3. Preserve locality for random writes.}
FDP reduces interference between sequential and random streams when random invalidations remain localized to maximize the use of the internal SSD erase blocks. With four sequential and four random streams, FDP reduces WAF from 3.49 to 1.59. 

\noindent\textbf{4. Separate hot and cold data.}
Placing hot random writes in a separate RUH prevents movement  of cold data during garbage collection and ensures their invalidations are localized. FDP maintains a WAF near 1 through 90\% prefill, compared with 1.81 without FDP.

\noindent\textbf{5. Separate streams with different access skew.}
Uniform and skewed random workloads should use different placement groups. Skewed workloads repeatedly invalidate a smaller set of LBAs, while uniform workloads distribute invalidations across a larger region. 

\noindent\textbf{6. Use the minimum number of placement groups that preserves distinct invalidation behavior.}
Assigning every file, tenant, or writer to a separate RUH is unnecessary and may exceed the limited number of RUHs exposed by the current generation of devices. Placement policies should first classify streams by invalidation rate, locality, working-set size and write rate, and then merge classes with similar behavior. 
\section{RocksDB Data Placement}
\label{sec:rocksdb}
\begin{comment}
\subsection{SST Lifetime Analysis}
To better understand the data lifetime characteristics of LSM file-levels to enable effective grouping by death times, we conducted an empirical study to analyze the actual lifetimes of SST files across different compaction levels. We instrumented RocksDB to track the creation and deletion timestamps of all SST files during a workload execution, enabling us to measure the actual persistence duration of files at each level.
\begin{figure}[t]
  \centering
  \includegraphics[width=0.9\columnwidth]{figures/rocksDB/lifetime_scatter_2x3.pdf}
 %\vspace{-2mm}
  \caption{SST file lifetime distribution across LSM levels (0–5)
  during the first 5 hours of compaction.}
  %\vspace{-3mm}
  \label{fig:lifetime_levels}
\end{figure}
Figure~\ref{fig:lifetime_levels} shows the distribution of file lifetimes across levels $L0$–$L5$ over a 5-hour workload. Lower levels ($L1$–$L3$) exhibit short, relatively uniform lifetimes, with most files persisting from seconds to minutes before compaction. From $L4$ onward, lifetime differences become evident, where $L4$ shows substantial variance (hundreds to tens of thousands of seconds), while $L5$ contains the oldest files, many persisting for the full duration. This clear lifetime stratification, emerging at $L4$, suggests these files should be grouped separately in the SSD rather than being mixed.
\end{comment}
In this section, we describe the data placement strategy used in RocksDB leveraging its LSM-tree-based architecture that only performs out of place writes on the storage layer. Based on our analysis of the distinct lifetimes of the WAL and SST levels (\S~\ref{sec:rocksdb-background}) and the isolation of concurrent sequential writes (\S~\ref{sec:seq-workloads}), we use four RUHs to separate the WAL, $L4$, $L5$ and higher, and $L0$–$L3$. We implement this data placement design through two storage paths: \\
1. Through the filesystem using \emph{experimental} VFS patches that propagate lifetime hints. (\S~\ref{sec:rocksdb_fdp_file_system}) \\
2. Through the raw character device using the \emph{stable} \texttt{io\_uring\_cmd} interface (\S~\ref{sec:torfs}).

\subsection{Data Placement Using Filesystem Hints}
\label{sec:rocksdb_fdp_file_system}
% \usepackage{booktabs}
% \usepackage{graphicx}
\begin{table}[]
\resizebox{\columnwidth}{!}{
\begin{tabular}{@{}lllll@{}}
\hline
Levels      & WAL   & L0$\sim$3 & L4   & \textgreater{}=L5 \\ \hline
Write Hints & SHORT & MEDIUM    & LONG & EXTREME           \\ \bottomrule
\hline
\end{tabular}
}
\caption{VFS write hint assignment for RocksDB.}
\label{tab:rocksdb_hints}
%\vspace{-3ex}
\end{table}
We modify RocksDB to exploit its architectural data separation through VFS write-lifetime hints. The WAL and SST levels already form distinct write streams with different expected lifetimes and write volumes, but this information is lost at the filesystem interface. When RocksDB creates a WAL or SST file, it sets a lifetime hint using \texttt{F\_SET\_FILE\_RW\_HINT}. RocksDB supports configurable data placement using the \texttt{fcntl()} system call. Its internal write lifetime hint enumeration (\texttt{WriteLifeTimeHint}) mirrors the Linux VFS \texttt{enum rw\_hint} values of \texttt{RWH\_WRITE\_LIFE\_}\{\texttt{SHORT,MEDIUM,LONG,EXTREME}\}.

We use the smallest number of lifetime classes supported by our architectural and empirical analysis (\S~\ref{sec:rocksdb-background}). The WAL is assigned \texttt{WLTH\_SHORT}, $L0$--$L3$ share \texttt{WLTH\_MEDIUM}, $L4$ uses \texttt{WLTH\_LONG}, and $L5$ and deeper levels share \texttt{WLTH\_EXTREME}. This policy uses four hints rather than assigning a separate hint to every level. It separates components only where their expected lifetimes differ substantially and avoids consuming additional RUHs. During flush and compaction, each output SST file receives the hint associated with its destination level. The SSD can therefore place writes with similar expected invalidation times in the same reclaim units, reducing data movement during garbage collection. \emph{The RocksDB integration requires about 25 lines of code changes.}

\subsection{Data Placement Using I/O Passthru (Filesystem Bypass) - \textbf{TorFS}}
\label{sec:torfs}
\begin{figure}[t]
\centering
\includegraphics[width=0.9\columnwidth]{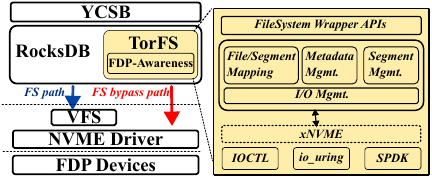}
\caption{Overview of FS and FS bypass on RocksDB.}
\label{fig:torfs}
%\vspace{-4ex}
\end{figure}
RocksDB provides abstract wrapper APIs for pluggable storage backends. \emph{To use FDP data placement with only stable Linux kernel interfaces, we developed TorFS (Figure~\ref{fig:torfs})}. TorFS implements the RocksDB \texttt{FileSystem} API and uses xNVMe~\cite{xnvme} to issue NVMe commands from user space through the \texttt{io\_uring\_cmd} interface utilizing the raw character device. It applies the data placement policy described in Table~\ref{tab:rocksdb_hints}. Instead of passing lifetime hints through the filesystem, TorFS directly specifies the placement identifier (PID) associated with the target RUH in each NVMe write command. This path bypasses the VFS, filesystem, and block layers while retaining the NVMe driver in the Linux kernel. TorFS consists of approximately 4,000 lines of C++ code. It provides a control path for evaluating whether bypassing the filesystem shows any performance advantage to control device WAF via application-controlled file to LBA mapping while preserving the same FDP placement policy and device WAF.

\minisec{FDP-Aware Data Placement}
TorFS intercepts \texttt{WriteLifeTimeHint} of  RocksDB and maps each file to one of the five FDP Reclaim Unit Handles (0--4). WAL files are mapped to RUH 0, $L0$--$L3$ SST files to RUH 1, $L4$ SST files to RUH 2, and $L5$+ SST files to  RUH 3. It additionally maps its MANIFEST files to RUH 4. This mapping groups data with similar lifetimes into physically isolated Reclaim Units.

\minisec{How Files Map Region on Raw Device} TorFS maintains a global mapping table to track the mapping between files and segments. Each segment defined by an offset and length, represents a range of LBAs. TorFS allocates one or more segments based on the file size and records the segments in the mapping table after files are flushed to the device, and uses the mapping table to quickly locate the segments and retrieve the data.

\minisec{Read and Write I/O Paths} Each file pre-allocates a DMA-pinned write buffer upon creation via the \texttt{xnvme\_buf\_alloc} API, avoiding repeated allocations on the critical path. 
Incoming writes are accumulated in this buffer. When the buffer is full or \texttt{Sync} is invoked, \texttt{DataSync()} aligns the payload to block boundaries, splits it according to the device’s maximum transfer size, and submits batched NVMe write commands tagged with the file’s Placement Handle.
TorFS issues aligned NVMe read commands via the \texttt{io\_uring\_cmd} I/O Passthru interface, bypassing the kernel page cache and eliminating kernel-to-user copies incurred by the VFS path.

\minisec{Metadata Management}
Metadata operations (e.g., file creation, update, rename, and deletion) are recorded as tagged, append-only entries in a dedicated \texttt{MetaSegment}. Each entry includes a typed header (\texttt{Base}, \texttt{Update}, \texttt{Replace}, \texttt{Delete}), avoiding in-place journaling overhead. When the log region of the MetaSegment reaches a threshold, a background thread asynchronously merges the accumulated incremental operations into a compact baseline snapshot and writes it back to the main metadata region. This compaction process does not block foreground write operations.

\minisec{Space Management and Deallocation}
TorFS mitigates post-deletion fragmentation using a priority-queue-based segment allocator. Free LBA ranges are maintained in a min-heap ordered by offset, enabling on-the-fly merging of contiguous segments and preserving allocation contiguity. Upon file deletion, segments are enqueued into a shared deallocate\_queue and processed by background worker threads. Workers merge adjacent segments and issue batched NVMe Dataset Management (\texttt{DSM}) Deallocate commands, informing the SSD of invalidated LBAs and reducing GC pressure.

To prevent overwhelming the SSD under delete-heavy workloads, TorFS employs a queue-depth-aware throttling policy with thresholds $T_{\text{low}}$ and $T_{\text{high}}$ ($T_{\text{low}} < T_{\text{high}}$). We use the following heuristic where $Q$ denotes the de-allocate queue depth:\\
If $Q < T_{\text{low}}$, all segments are reclaimed immediately.\\
If $T_{\text{low}} \leq Q < T_{\text{high}}$, only segments larger than 64 KB are reclaimed.\\
If $Q \geq T_{\text{high}}$, only segments larger than 1 MB are reclaimed.\\
This policy ensures that high delete pressure does not flood the SSD controller with fine-grained Deallocate commands.
\section{MySQL Data Placement}
\label{sec:mysql}
We modify MySQL to expose the architectural separation among its major write streams through VFS lifetime hints. InnoDB generates three distinct heavy write patterns due to redo-log writes, double-write buffer writes, and buffer-manager writes to user tablespaces. The redo log is written sequentially over a large cyclic region. The double-write buffer is also written sequentially but is reused more frequently because it occupies a smaller cyclic region. Buffer-manager flushes produce random, in-place writes over much larger user tablespaces. 

When MySQL opens each file, it assigns a lifetime hint using \texttt{F\_SET\_FILE\_RW\_HINT}. We assign \texttt{WRITE\_LIFE\_SHORT} to double-write buffer files, \texttt{WRITE\_LIFE\_MEDIUM} to redo-log files, and \texttt{WRITE\_\allowbreak LIFE\_\allowbreak EXTREME} to user tablespaces written by the buffer manager. The double-write buffer receives the "short" hint because its small cyclic region is overwritten frequently. The larger redo-log region is reused less frequently and receives a "medium" hint. User tablespaces receive the "extreme" hint because their pages are distributed over a large address space and generally remain valid longer. Temporary files are set to "long" while metadata files to store schema and control information are set to "not set". The VFS hints are just used to separate data and their relative temperature is not semantically important.

\textbf{Why only VFS integration?} An I/O Passthru implementation similar to TorFS (\S~\ref{sec:torfs}) was impractical within the scope of this paper. RocksDB exposes a pluggable \texttt{FileSystem} API and primarily writes immutable SST files, making a character device practical. InnoDB instead relies on filesystem semantics for mutable user tablespaces, redo logs, temporary files, and double-write buffers. Our RocksDB evaluation (\S~\ref{sec:rocksdb-eval}) already compares filesystem hinting with filesystem bypass and isolates their effects on WAF and performance. 

\section{Evaluation}
\label{sec:evaluation}
%\subsection{Experimental Setup}

Our evaluation covers RocksDB and MySQL. All experiments are conducted on a single server equipped with dual Intel Xeon Gold 6342 processors (2$\times$24 cores, 2.80\,GHz), 498\,GB DRAM, and a 3.84\,TB FDP-capable NVMe SSD, running Linux 6.9.0. We use RocksDB 7.2.2 and Percona Server 8.0.34. For RocksDB, we compare three configurations, namely TorFS with FDP, FDP on XFS (\S~\ref{sec:rocksdb_fdp_file_system}), and a conventional baseline on XFS, using YCSB~\cite{ycsb} and \texttt{db\_bench}~\cite{rocksdb} workloads. For MySQL, we run InnoDB on ext4 using TPC-C to evaluate the data separation benefits outlined in \S~\ref{sec:mysql}.

% We should ideally identify the claims of the evaluation under each of these sections. Ideally some of those should be what is insightful for that domain.
% A specific tuning point might not be insightful for that application but rather a larger design pattern that helps the application would be more scientifically defensible.

% It will help if we identify the claims in each of these sections. 
%\subsection{RocksDB}
% Using application benchmarks for key value stores e.g., YCSB
% Are there any benchmarks that use AI workloads with key value stores? 

% Yahoo! Cloud Serving Benchmark (YCSB) is a standardized industry-level macro-benchmarking framework developed by Yahoo, widely utilized for comparative performance evaluation of diverse databases, while simulating typical workloads observed in real-world application scenarios \cite{ycsb}. We evaluated our system using \texttt{db\_bench} and YCSB workloads running on RocksDB, covering both single-workload and multi-tenant execution modes. In the single-workload setting, each \texttt{db\_bench} and YCSB workload was executed independently to characterize baseline performance—write amplification, throughput and latency. In the multi-tenant setting, multiple YCSB workloads were issued concurrently to emulate realistic deployment scenarios with competing tenants. This design allows us to assess performance isolation, resource contention, and the impact on write amplification under concurrent accesses.

\subsection{RocksDB}
\label{sec:rocksdb-eval}
In this section, we evaluate RocksDB using multiple workloads from YCSB for macro-benchmarks simulating realistic cloud applications, and \texttt{db\_bench} for
microbenchmarks targeting specific I/O patterns used in past SSD interface work~\cite{zns:BjorlingAHRMGA21}. Our experiments cover single-tenant and multi-tenant settings. In the single-tenant setting, each benchmark executes independently to characterize baseline write amplification, throughput, and latency.
In the multi-tenant setting, multiple instances of the workload and RocksDB are run concurrently to emulate realistic deployments with competing tenants, to assess performance isolation, resource contention, and write amplification under concurrent tenant access.

\subsubsection{YCSB-A (Update-Intensive)}
\label{sec:evaluation-update}
We evaluate FDP's performance under YCSB Workload~A using the standard read-update ratio (50/50) and a highly write-intensive ratio (20/80) to stress the system, comparing FDP on XFS and TorFS against a Conv baseline.
Both configurations use 40 million records, each consisting of a default YCSB key ($\sim$23\,B) and a 40\,KB value (10 fields of 4\,KB each),  totaling approximately 1.6\,TB, together with 200 million operations, 6 client threads, and the standard request distribution with a Zipfian constant of 0.99. The total amount of data written to the SSD over the full experiment is approximately 94\,TB.

\begin{figure}[t]
    \centering
    \begin{subfigure}{0.48\linewidth}
        \centering
        \includegraphics[width=\linewidth]{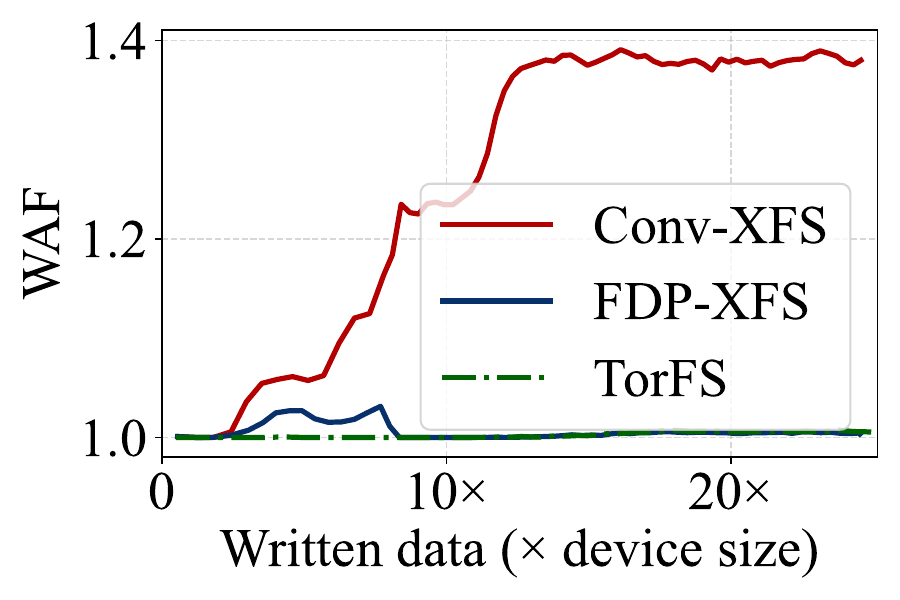}
        \caption{50\% R / 50\% U}
        \label{fig:waf_single_5050}
    \end{subfigure}
    \hfill
    \begin{subfigure}{0.48\linewidth}
        \centering
        \includegraphics[width=\linewidth]{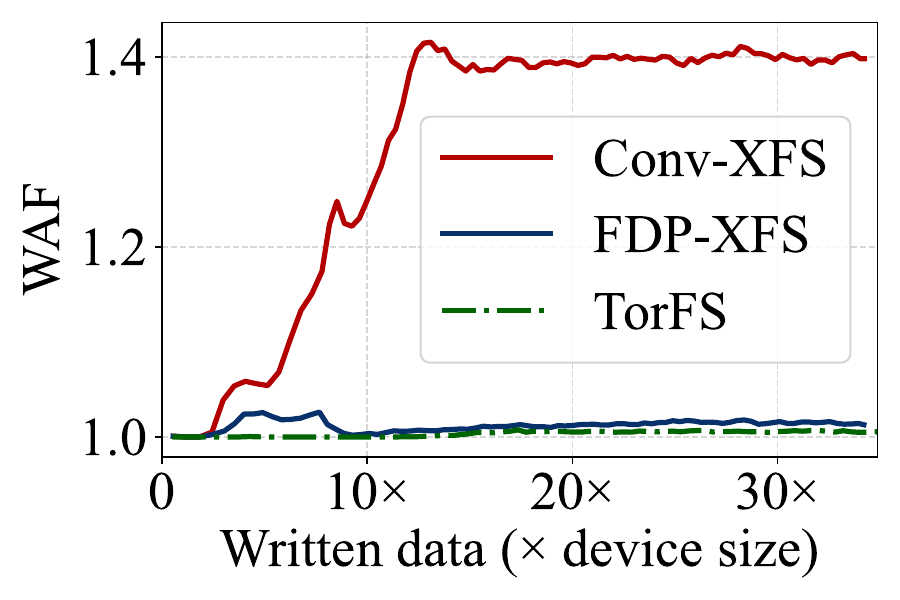}
        \caption{20\% R / 80\% U}
        \label{fig:waf_single_2080}
    \end{subfigure}
    \caption{WAF  under different read/update ratios for YCSB Workload~A.}
    \label{fig:waf_single}
       %\vspace{-4ex}
\end{figure}

Figure~\ref{fig:waf_single} shows the result of this experiment. Both FDP-XFS and TorFS reach an ideal WAF of \textasciitilde1 across both ratios after a brief warm-up, while the Conv baseline rises to approximately 1.39 with noticeable spikes. The near-perfect overlap between TorFS and FDP-XFS confirms that FDP's benefit stems from its placement isolation mechanism rather than TorFS-specific (OS bypass) optimizations. The consistent results across both read-update ratios demonstrate that FDP's data separation effectiveness is robust to increasing update intensity and write rates under out-of-place writes (delete-append write pattern).

\begin{figure}[t]
    \centering
    \begin{subfigure}{0.48\linewidth}
        \centering
        \includegraphics[width=\linewidth]{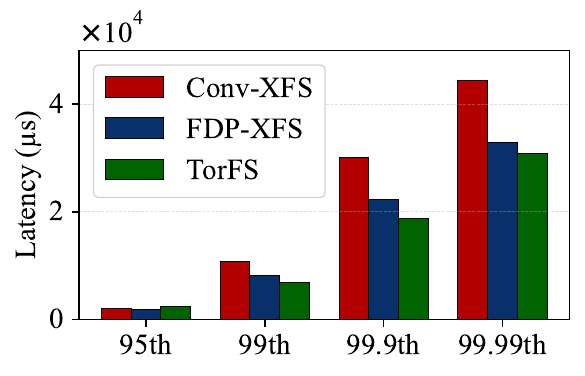}
        \caption{Read latency}
        \label{fig:qos_read_5050}
    \end{subfigure}
    \hfill
    \begin{subfigure}{0.48\linewidth}
        \centering
        \includegraphics[width=\linewidth]{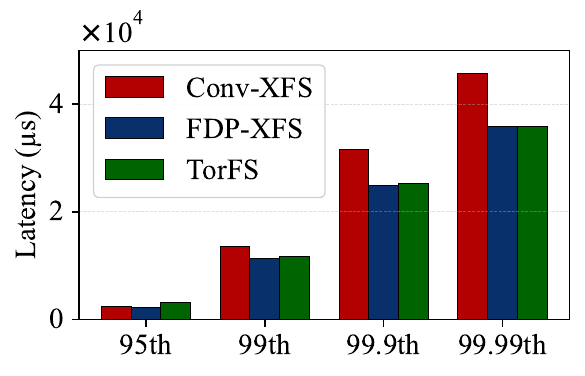}
        \caption{Update latency}
        \label{fig:qos_update_5050}
    \end{subfigure}
    \caption{Tail latency percentiles for YCSB Workload~A (50/50 read-update).}
    \label{fig:qos_5050}
\end{figure}

Figure~\ref{fig:qos_5050} shows tail latency at 95th, 99th, 99.9th, and 99.99th  percentiles. FDP variants show lower tail latencies than Conv due to less GC interference. FDP-XFS reduces p99.99 read and update latency from 44\,383\,µs and 45\,791\,µs (Conv) to 32\,959\,µs and 35\,775\,µs. TorFS achieves the maximum gain by minimizing OS layer overhead, with p99.99 read and update latencies of 30\,815\,µs and 35\,775\,µs, representing reductions of 31\% and 22\% over Conv. The results for the 20/80 workload are provided in\conditionaltext{ the extended paper in supplementary materials}{ \S~\ref{app:update_qos_2080}(supplementary material)} and show consistent behavior, demonstrating the effectiveness of our data separation policies against write amplification in LSM-tree-based key-value stores while preserving QoS for latency-sensitive workloads.

\vspace{-7pt}
\subsubsection{Modified YCSB-E (Insert-Intensive)}
We evaluate the robustness of our data separation policy using YCSB Workload~E consisting of 95\% inserts and 5\% scans, simulating data ingestion patterns typical of logging systems, time-series databases, and event streaming applications. The workload inserts 90 million records in total, occupying approximately 3.4\,TB (95\% of device capacity). 

\begin{figure}[t]
    \centering
    \begin{subfigure}{0.45\linewidth}
        \centering
        \includegraphics[width=\linewidth]{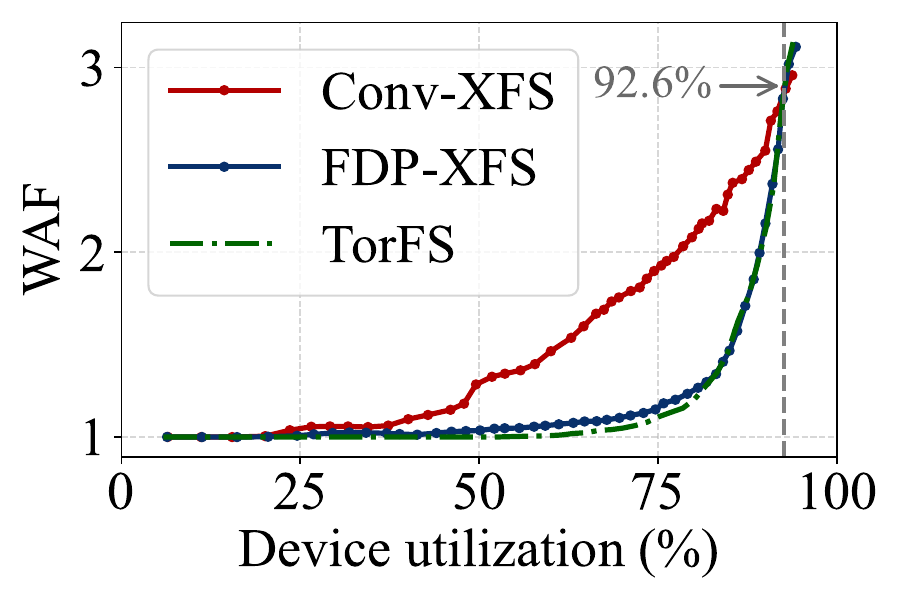}
        \caption{WAF vs utilization.}
        \label{fig:waf_insert}
    \end{subfigure}
    \hfill
    \begin{subfigure}{0.5\linewidth}
        \centering
        \includegraphics[width=\linewidth]{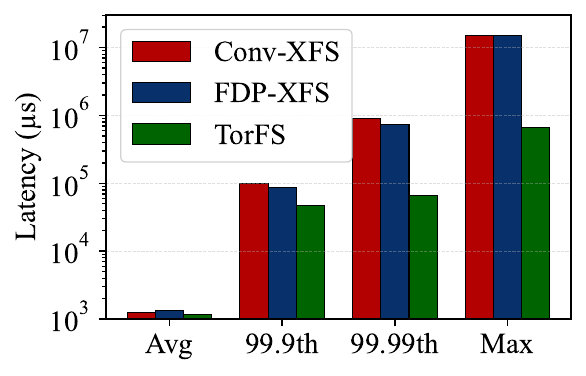}
        \caption{Insert latency percentiles.}
        \label{fig:qos_insert}
    \end{subfigure}
\caption{WAF and tail latency under modified YCSB Workload~E.}
\label{fig:insert_intensive}
  %\vspace{-4ex}
\end{figure}

Figure~\ref{fig:waf_insert} plots interval WAF against device utilization as the insert-heavy workload grows the dataset from roughly 6\% to 94\% of the device capacity. Conv-XFS diverges early because data mixing from SST and WAL files during compaction, as new inserts trigger compactions that repeatedly co-locate and rewrite data from different insertion phases, pushes its WAF above 1.10 at about 43\% utilization and to 2.09 at 80\%. Both FDP-XFS and TorFS suppress this effect substantially. FDP-XFS stays below 1.10 until roughly 70\% utilization and reaches only 1.25 at 80\%, while TorFS remains near 1.00 up to 60\% and reaches 1.20 at 80\%. The gap is widest between 70\% and 85\% utilization, where Conv-XFS incurs 1.6--1.7$\times$ the WAF of either FDP configuration. Beyond roughly 85\%, however, all three configurations degrade sharply as the free space available to garbage collection vanishes, and the FDP curves meet Conv-XFS at about 92.6\% utilization (dashed line), where all three reach a WAF of about 2.9. Placement isolation is therefore most beneficial for insert-heavy workloads while the device retains free space. As the device approaches full capacity, GC pressure from the shortage of free blocks dominates and placement alone can no longer contain write amplification.

Figure~\ref{fig:qos_insert} shows insert tail latency across four percentile tiers. The three approaches exhibit broadly comparable latency at p95 through p99.9. \emph{At p99.99, TorFS achieves dramatically lower latency than both FDP-XFS and Conv-XFS (106,623~µs vs. 2,209,791~µs and 2,529,279~µs, a 20.7$\times$ and 23.7$\times$ reduction). This extreme-tail advantage suggests TorFS's direct device access and deallocation policies reduce rare I/O queuing and GC overheads substantially, delivering superior QoS precisely where latency spikes are most disruptive to application responsiveness.}

\subsubsection{SSD Utilization Stress Test}
To evaluate FDP's effectiveness under sustained high SSD utilization conditions, we conduct a stress test using YCSB Workload~A (50/50 read-update) with 75 million records totalling approximately 3\,TB. As the workload progresses, continuous updates drive device utilization above 90\%, allowing us to examine how FDP's WAF benefit evolves as available SSD space nears exhaustion. We could reduce this value by increasing the compaction frequency but we retain the default setting to stress device utilization.

\begin{figure}[t]
    \centering
    \begin{subfigure}{0.48\linewidth}
        \centering
        \includegraphics[width=\linewidth]{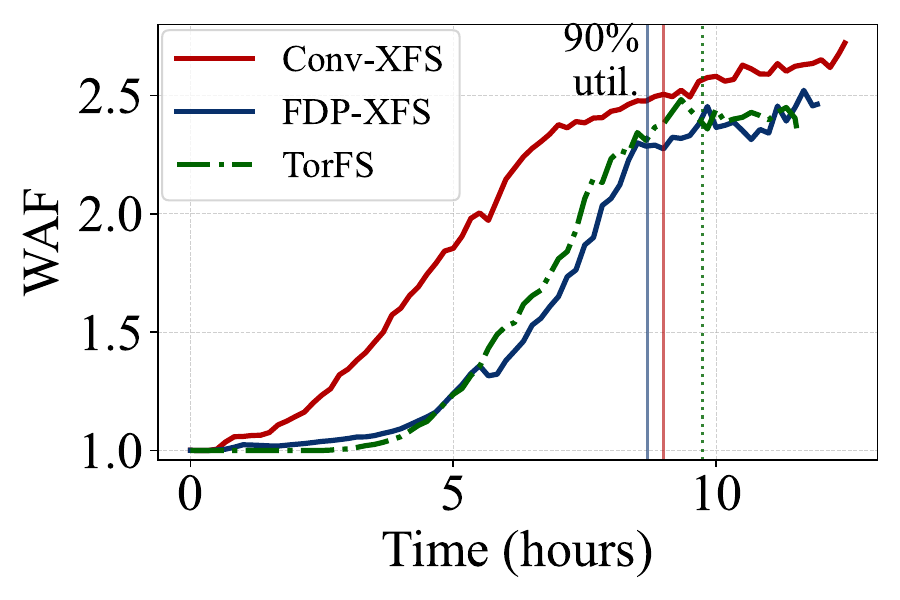}
        \caption{WAF over time. }
        \label{fig:waf_large_vs_time}
    \end{subfigure}
    \hfill
    \begin{subfigure}{0.48\linewidth}
        \centering
        \includegraphics[width=\linewidth]{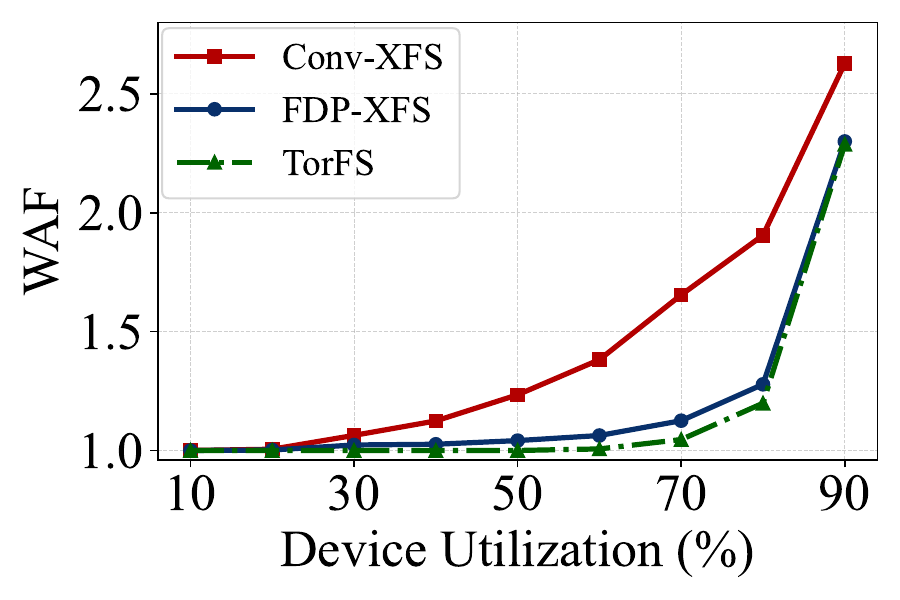}
        \caption{WAF vs.\ device utilization.}
        \label{fig:waf_large_util}
    \end{subfigure}
    \caption{WAF comparison under large-scale stress test.}
      %\vspace{-6pt}
    \label{fig:waf_large}
\end{figure}

Figure~\ref{fig:waf_large} presents two complementary views of the same experiment, WAF over time and WAF at each SSD utilization level, respectively.
Below 30\% utilization, all three configurations maintain WAF near 1.0 with negligible differences. FDP's advantage grows steadily as utilization increases, peaking between 70\% and 80\%, where FDP-XFS and TorFS reduce WAF by 33\% and 37\% over Conv-XFS (from 1.91 to 1.28 and 1.20, respectively). This is the region in which lifetime-based data isolation is most effective with sufficient Reclaim Unit capacity remaining to sustain clean per-RU GC boundaries. Beyond 90\% utilization (vertical lines at approximately 8.7, 9.0, and 10.4 experiment hours for FDP-XFS, Conv-XFS, and TorFS), the SSD OP space cannot be effectively used to buffer the writes and it is forced to co-locate data with different lifetime hints within the same Reclaim Units, and the gap narrows to 12--13\% (Conv: 2.63, FDP-XFS: 2.30, TorFS: 2.29). Nevertheless, FDP consistently outperforms Conv across all utilization levels, confirming that lifetime-aware placement provides WAF benefits even at high practical SSD utilizations. The latency results from the same experiment follow a similar pattern, with p95 through p99.9 remaining broadly comparable across configurations while p99.99 shows meaningful reductions under FDP, as detailed
in\conditionaltext{ the extended paper in supplementary materials.}{ \S~\ref{app:ssd_uti_stress_latency} (supplementary material).}

% To evaluate FDP's performance under more demanding conditions, we
% conduct a stress test with significantly larger datasets that push
% device utilization above 90\%, examining FDP's effectiveness when
% storage capacity pressure increases compaction frequency and intensity.

% Figure~\ref{fig:waf_large_vs_size} tracks WAF evolution throughout a
% Workload~A run, with the x-axis representing cumulative written volume
% as a multiple of device capacity. As written volume increases, Conv-XFS
% WAF rises steadily, reaching 1.84 at 15$\times$ device capacity.
% FDP-XFS and TorFS remain substantially lower throughout, reaching 1.24
% and 1.06 respectively at the end of the run, achieving 33\% and 43\%
% WAF reduction relative to Conv-XFS.

% Figure~\ref{fig:waf_large_util} shows WAF as a function of device
% utilization at different points during the same run. Below 40\%
% utilization, all three systems maintain WAF near 1.0--1.1. As
% utilization climbs past 60\%, Conv-XFS escalates noticeably, reaching
% approximately 1.6 at 80\% and 2.5 at 90\%, while FDP-XFS and TorFS
% remain stable through 70\% utilization before rising sharply as free
% space becomes critically scarce.

% Figure~\ref{fig:qos_large} shows the tail latency distribution under
% these conditions. At the 95th and 99th percentiles, all configurations
% exhibit comparable latency. At the 99.99th percentile, FDP-XFS reduces
% read and update latency by 26.9\% and 20.7\% respectively over
% Conv-XFS, while TorFS achieves 25.9\% and 18.4\% reductions.

% \begin{figure}[t]
%     \centering
%     \begin{subfigure}{0.48\linewidth}
%         \centering
%         \includegraphics[width=\linewidth]{figures/evaluation/rocksdb/single_large/qos_read_percentiles_60m_200m.pdf}
%         \caption{Read latency}
%         \label{fig:qos_read_large}
%     \end{subfigure}
%     \hfill
%     \begin{subfigure}{0.48\linewidth}
%         \centering
%         \includegraphics[width=\linewidth]{figures/evaluation/rocksdb/single_large/qos_update_percentiles_60m_200m.pdf}
%         \caption{Update latency}
%         \label{fig:qos_update_large}
%     \end{subfigure}
%         \caption{Tail latency under large-scale stress test.}
%        %\vspace{-6pt}
%     \label{fig:qos_large}
% \end{figure}

\vspace{-6pt}
\subsubsection{db\_bench Workloads}

% We evaluate FDP's write amplification behavior under two representative \texttt{db\_bench} workloads: \texttt{fillrandom}, which models bulk data ingestion via sequential random-key inserts, and \texttt{overwrite}, which models steady-state update pressure via repeated overwrites of existing keys. We compare FDP-enabled configurations on both XFS and our prototype TorFS against a baseline without FDP on Conv. Figure~\ref{fig:waf_dbbench} demonstrates that FDP consistently maintains low write amplification across both workload configurations.

We evaluate RocksDB's WAF under two representative \texttt{db\_bench} workloads used in previous ZNS studies~\cite{zns:BjorlingAHRMGA21}, \texttt{fillrandom}, which models bulk data ingestion via sequential random-key inserts, and \texttt{overwrite}, which models steady-state update pressure via repeated overwrites of existing keys.
We compare FDP-enabled configurations on both XFS and TorFS against a Conv baseline on XFS.
Both workloads target approximately 2\,TB of device capacity. The \texttt{overwrite} workload uses a two-phase procedure: an initial \texttt{fillseq} phase rapidly populates the full key space, followed by an overwrite phase that drives sustained compaction pressure during the measurement period.

\begin{figure}[t]
    \centering
    \begin{subfigure}{0.48\linewidth}
        \centering
        \includegraphics[width=\linewidth]{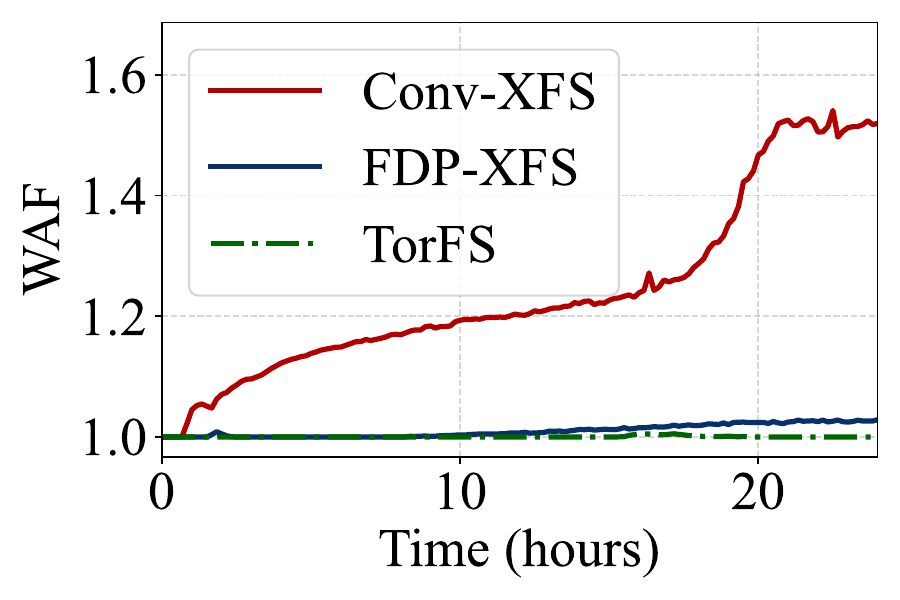}
        \caption{\texttt{fillrandom}}
        \label{fig:waf_fillrandom}
    \end{subfigure}
    \hfill
    \begin{subfigure}{0.48\linewidth}
        \centering
        \includegraphics[width=\linewidth]{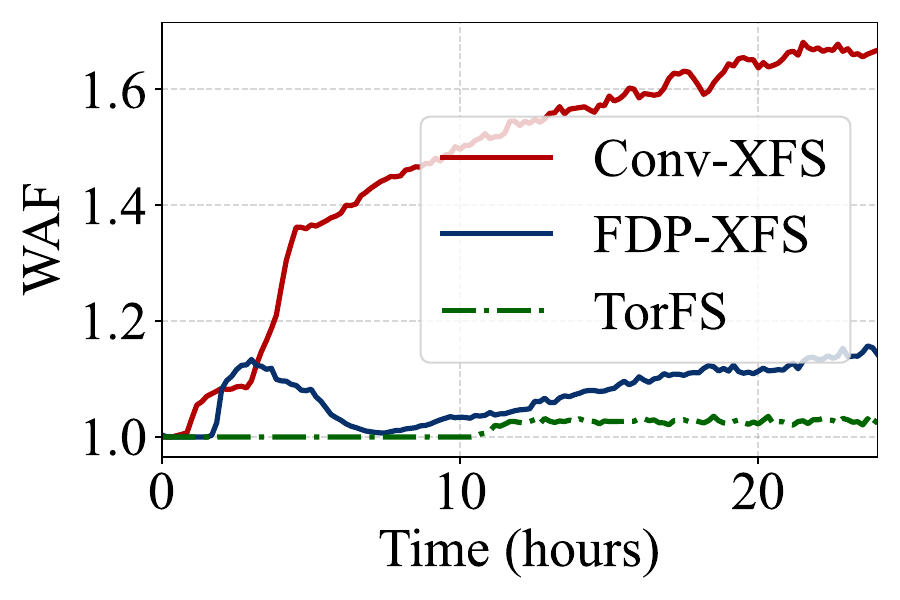}
        \caption{\texttt{overwrite}}
        \label{fig:waf_overwrite}
    \end{subfigure}
    \caption{WAF under \texttt{db\_bench} workloads.}
    %\vspace{-3ex}
    \label{fig:waf_dbbench}
\end{figure}

Under the \texttt{fillrandom} workload (Figure~\ref{fig:waf_fillrandom}), Conv-XFS exhibits rapid WAF escalation as the dataset grows and compaction activity intensifies, reaching approximately 1.5--1.6. FDP-XFS achieves moderate improvement by directing data to separate reclaim units, while TorFS maintains a near-flat WAF of approximately 1.05 throughout the experiment. Under the \texttt{overwrite} workload (Figure~\ref{fig:waf_overwrite}), the divergence between configurations becomes more pronounced over time. Conv-XFS again escalates significantly, reflecting continuous invalidation of stale data across mixed-temperature blocks, while TorFS sustains a stable WAF near 1.05 for the duration of the workload. The advantage of TorFS over FDP-XFS across both workloads suggests that filesystem-level co-design with FDP is important to fully realizing the device's placement capabilities.

\subsubsection{Filesystem Sensitivity on RocksDB} 

\label{sec:rocksdb_fs_sensitivity}

% \begin{figure}[t]
%   \centering
%   \includegraphics[width=0.8\linewidth]{figures/evaluation/rocksdb/fs_sensi/ext4_vs_xfs.pdf}
%   \caption{ WAF over the full RocksDB test run, Conv vs.\ FDP on ext4 and xfs.}
%   \label{fig:rocksdb_fs_waf}
% \end{figure}

\begin{figure}[t]
  \centering
  \begin{minipage}[t]{0.45\linewidth}
    \centering
    \includegraphics[width=\linewidth]{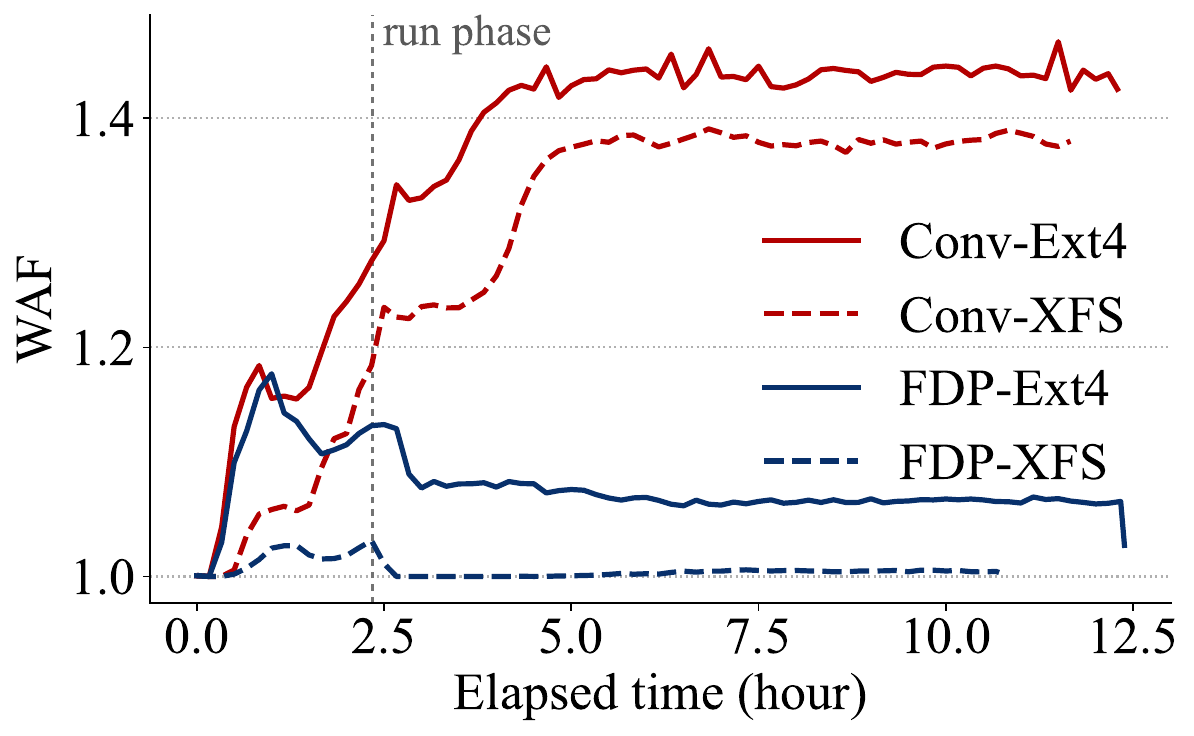}
    \captionof{figure}{WAF under Conv and FDP on ext4 and XFS.}
    \label{fig:rocksdb_fs_waf}
  \end{minipage}
  \hfill
  \begin{minipage}[t]{0.45\linewidth}
    \centering
    \includegraphics[width=\linewidth]{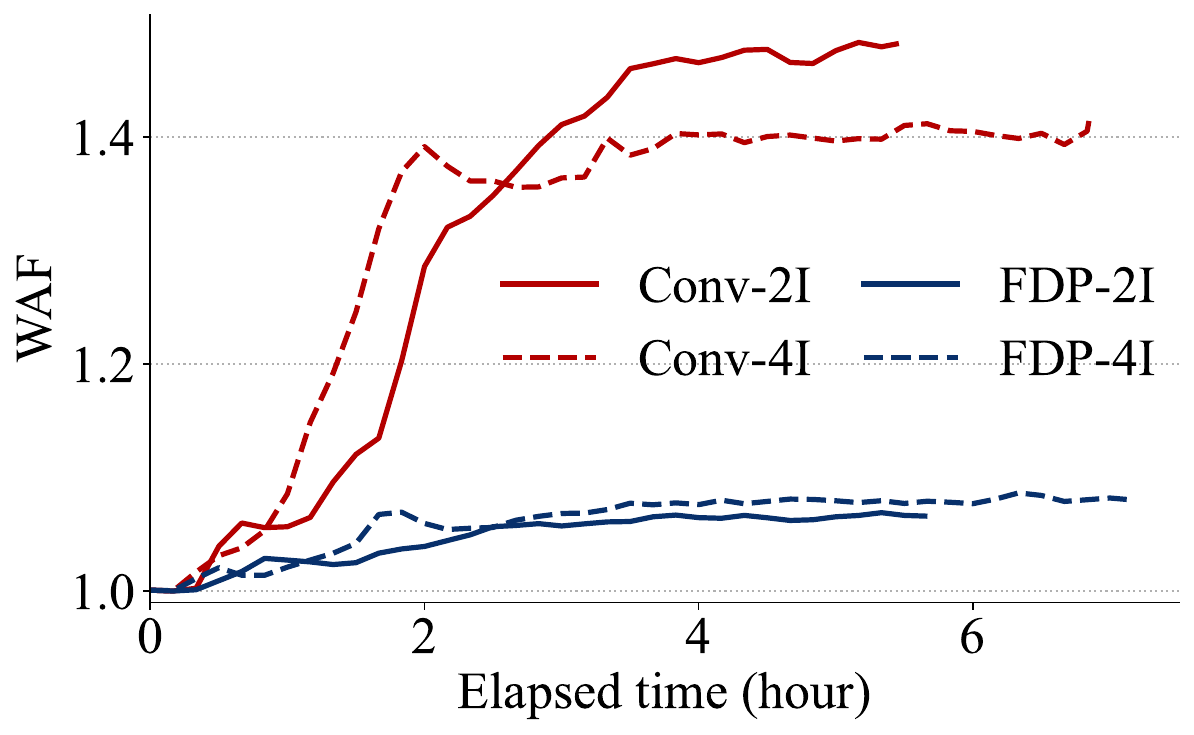}
    \captionof{figure}{ WAF for 2- and 4-instance RocksDB.}
    \label{fig:multi_rocksdb_waf}
  \end{minipage}
\end{figure}

{Figure~\ref{fig:rocksdb_fs_waf} shows that FDP's WAF benefit on RocksDB is essentially filesystem-neutral. Both curves peak early and decay under FDP, instead of climbing monotonically as under Conv, and both settle far below their Conv baseline, reducing steady state WAF by 26.0\% on ext4 (1.44 to 1.07) and 27.3\% on XFS (1.38 to 1.00). \emph{The one gap that does not close is between the filesystems themselves. ext4 and XFS differ by about 0.06 WAF under both Conv and FDP, and under FDP that gap makes up most of ext4's residual overhead above the ideal value of~1 (0.066 for ext4 versus 0.005 for XFS), pointing to ext4's own allocation and journaling behaviour rather than to placement. Because XFS consistently reaches the lower WAF of the two, we use it as the baseline filesystem for the remaining RocksDB experiments.}}

\subsubsection{Multi-Tenant RocksDB Deployment}
Multi-tenant deployments with containerized RocksDB instances represent a common pattern in modern cloud infrastructure. To evaluate FDP's
effectiveness in such scenarios, we deploy two and four containerized RocksDB instances executing concurrent YCSB Workload~A (50/50 read-update) sharing the same FDP SSD while scaling the setup. Each container maps RocksDB levels to placement handles as described in Table~\ref{tab:rocksdb_hints} and shares the RUHs across the instances for data of similar lifetimes. Both configurations maintain a consistent aggregate dataset of approximately 1.6\,TB. The two-instance setup allocates 20 million records per instance, while the four-instance setup allocates 10 million records per instance, each issuing 50 million operations.

% \begin{figure}[t]
%     \centering
%     \includegraphics[width=0.9\linewidth]{figures/evaluation/rocksdb/multi-tenants/multi_instances_waf_lat.pdf}
%     \caption{WAF over time and average latency for 2- and 4-instance RocksDB configurations. Solid lines/bars: 2-instance; dashed lines/hatched bars: 4-instance. }
%     \label{fig:multi_rocksdb}
%    %\vspace{-3ex}
% \end{figure}

Figure~\ref{fig:multi_rocksdb_waf} shows WAF and latency results under multi-tenant workloads. FDP achieves steady-state WAF of 1.07 and 1.08 for 2- and 4-instance configurations, compared to 1.48 and 1.40 for Conv, representing 28\% and 23\% reductions. This improvement stems from FDP's isolation of data of similar lifetimes in RU's preventing unnecessary data movement. Since RUHs are shared across containers, data of similar lifetimes from different containers are mapped to the same RUH, preventing intermixing across tenants and leading to more efficient use of SSD internal OP. Average read and update latencies remain comparable across both configurations, indicating that FDP's placement mechanism does not introduce measurable throughput overhead under multi-tenant workloads. These results highlight the benefit of sharing RUHs across tenants with the same placement policy. Although the number of RUHs on an SSD is limited and tenants must share them rather than each owning a dedicated set, FDP remains effective across multi-tenant configurations.
\subsection{MySQL}

We evaluate FDP on Percona Server 8.0.34 with InnoDB on ext4 using \texttt{tpcc-mysql}'s TPC-C workload~\cite{tpcc_mysql} with 128 concurrent client connections to sustain SSD write pressure. Each experiment comprises a \emph{data loading phase} that populates the warehouses, followed by a \emph{transaction-processing phase} that executes five transaction types (New-Order, Payment, Order-Status, Delivery, and Stock-Level)~\cite{leutenegger1993modeling}.

% We evaluate FDP's effectiveness on Percona Server 8.0 with InnoDB under EXT4, using the industry standard OLTP benchmark (TPC-C) benchmark. TPC-C consists of a \emph{data loading phase} that initialises a set of warehouses to establish a realistic data distribution, followed by a \emph{stress-testing phase} that drives a mixed workload of five transaction types (New-Order, Payment, Order-Status, Delivery, Stock-Level) under high concurrency~\cite{leutenegger1993modeling}.

\subsubsection{File Classification Strategy}
This section evaluates how assigning different Reclaim Unit Handles (RUHs) to different file types affects WAF. Based on the write patterns and relative write volumes of each file category described in \S~\ref{sec:mysql}, we define five placement policies and assess their WAF impact under a TPC-C workload of 6000
warehouses. WAF values are sampled at multiple time points during the final hour of each run and averaged to obtain a representative steady-state metric.
The five policies and their resulting average WAF are summarised in Table~\ref{tab:mysql_policy}.

\begin{table}[htbp]
\centering
\caption{File placement policies and average WAF.}
%\vspace{-3ex}
\label{tab:mysql_policy}
\setlength{\tabcolsep}{4pt}
\small
\begin{threeparttable}
\begin{tabular}{@{}ccccccc@{}}
\toprule
\textbf{Policy} &
\textbf{User Table} &
\textbf{DW Buffer} &
\textbf{Log} &
\textbf{Temp} &
\textbf{Meta} &
\textbf{WAF} \\
\midrule
1 & \texttt{EXT} & \texttt{SHT} & \texttt{MED} & \texttt{LNG} & \texttt{NS} & 1.24 \\
2 & \texttt{EXT} & \texttt{SHT} & \texttt{MED} & \texttt{MED} & \texttt{NS} & 1.24 \\
3 & \texttt{EXT} & \texttt{SHT} & \texttt{NS}  & \texttt{NS}  & \texttt{NS} & 1.24 \\
4 & \texttt{EXT} & \texttt{NS}  & \texttt{NS}  & \texttt{NS}  & \texttt{NS} & 1.53 \\
5 (Conv.) & \texttt{NS}  & \texttt{NS}  & \texttt{NS}  & \texttt{NS}  & \texttt{NS} & 1.80\\
\bottomrule
\end{tabular}

\begin{tablenotes}[para,flushleft]
\footnotesize
\textit{DW Buffer} = Double-Write Buffer.
\texttt{EXT} = WLTH\_EXTREME,
\texttt{SHT} = WLTH\_SHORT,
\texttt{MED} = WLTH\_MEDIUM,
\texttt{LNG} = WLTH\_LONG,
\texttt{NS} = WLTH\_NOT\_SET.
\end{tablenotes}
%\vspace{-2ex}

\end{threeparttable}
\end{table}

% \begin{table}[t]
% \centering
% \caption{\bluetext{Host write volume by InnoDB file category}}
% \label{tab:mysql_ruh_breakdown}
% \begin{tabular}{@{}lr@{}}
% \toprule
% File category & Share of host writes \\
% \midrule
% User Table  & 44.9\% \\
% DW Buffer   & 43.7\% \\
% Log         & 4.3\%  \\
% Temp        & $<$0.1\% \\
% Meta & 7.0\% \\
% \bottomrule
% \end{tabular}
% \begin{flushleft}
% \end{flushleft}
% \end{table}

Table~\ref{tab:mysql_policy} yields three observations. First, Policies~1--3 all achieve a WAF of 1.24 at the reported precision, indicating that assigning log, temporary table, and metadata files to additional RUHs provides no observable WAF benefit. This result is consistent with their similar lifetimes and small share of the total write volume. Second, Policy~4 reduces WAF from 1.80 to 1.53 (15.0\%) relative to Policy~5 (Conv) by isolating user table files, suggesting that separating long-lived user data mitigates GC interference among files with different lifetimes. Third, Policy~3 further reduces WAF from 1.53 to 1.24 (19.0\%) relative to Policy~4 by separating double-write buffer files from user table files. This additional improvement is consistent with the high write volume and frequent overwrites of double-write buffers. 
Policy~3 achieves the same 31.1\% WAF reduction relative to Conv as Policies~1 and~2 while using only three RUHs, the fewest among these policies. It maps user table files to \texttt{EXTREME}, double-write buffer files to \texttt{SHORT}, and all remaining files to the default \texttt{NOT\_SET}. We therefore adopt Policy~3 as the baseline placement policy for all subsequent experiments.

\subsubsection{Scalability Across SSD Utilization} 

Building on the minimal-RUH policy established above, we further evaluate FDP's effectiveness as SSD utilization increases within a single sustained workload. We run TPC-C under both Conv and FDP at two warehouse counts, 6000 (15 hours) and 12000 (6 hours), letting device utilization climb from near-empty toward 90\% over the course of each run, and report WAF as a function of the utilization level reached.
\begin{figure}[t]
    \centering
    \begin{minipage}{0.45\linewidth}
        \centering
        \includegraphics[width=\linewidth]{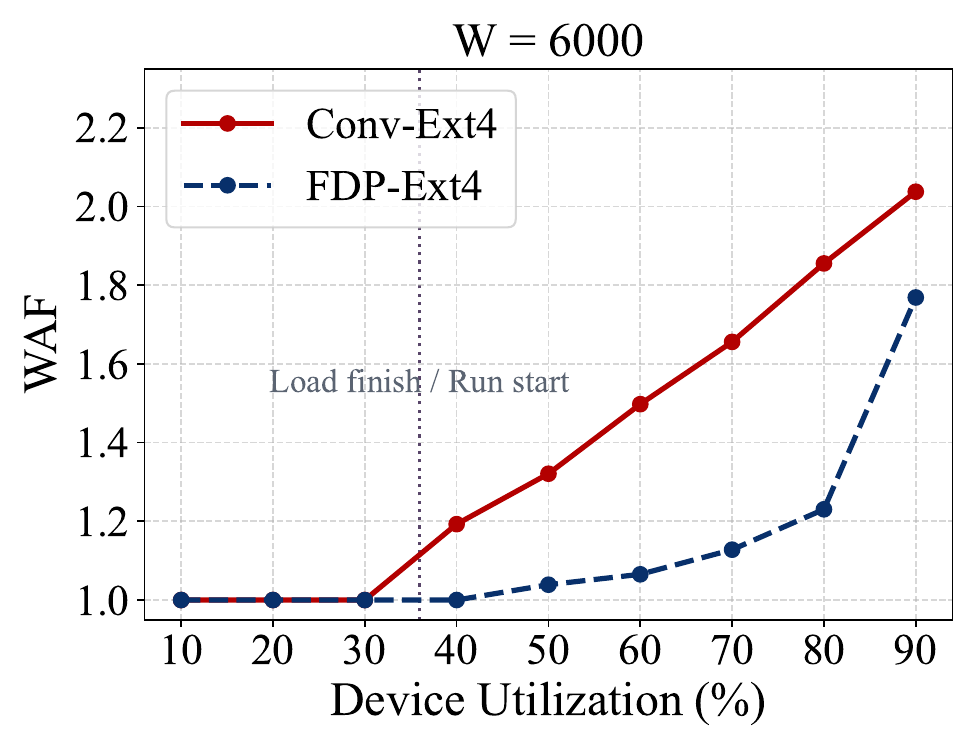}
    \end{minipage}
    \hfill
    \begin{minipage}{0.45\linewidth}
        \centering
        \includegraphics[width=\linewidth]{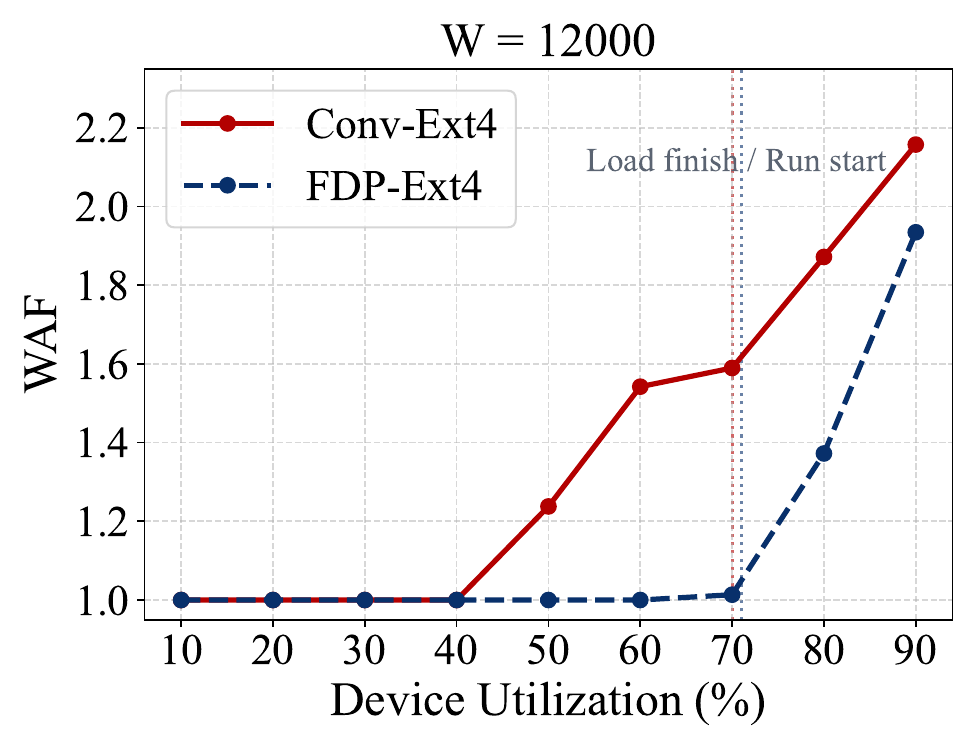}
    \end{minipage}
         \caption{{WAF as device utilization increases during TPC-C runs with 6000 (left) and 12000 (right) warehouses.}}

    \label{fig:waf_warehouse}
\end{figure}

\begin{figure}[t]
    \centering
    \begin{minipage}{0.45\linewidth}
        \centering
        \includegraphics[width=\linewidth]{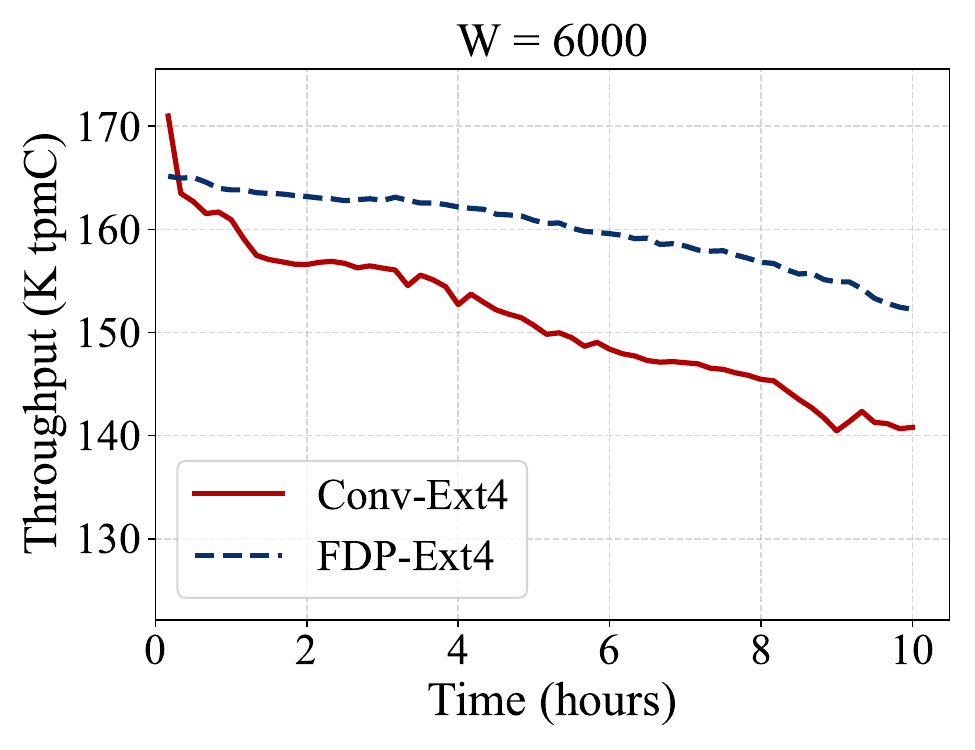}
    \end{minipage}
    \hfill
    \begin{minipage}{0.45\linewidth}
        \centering
        \includegraphics[width=\linewidth]{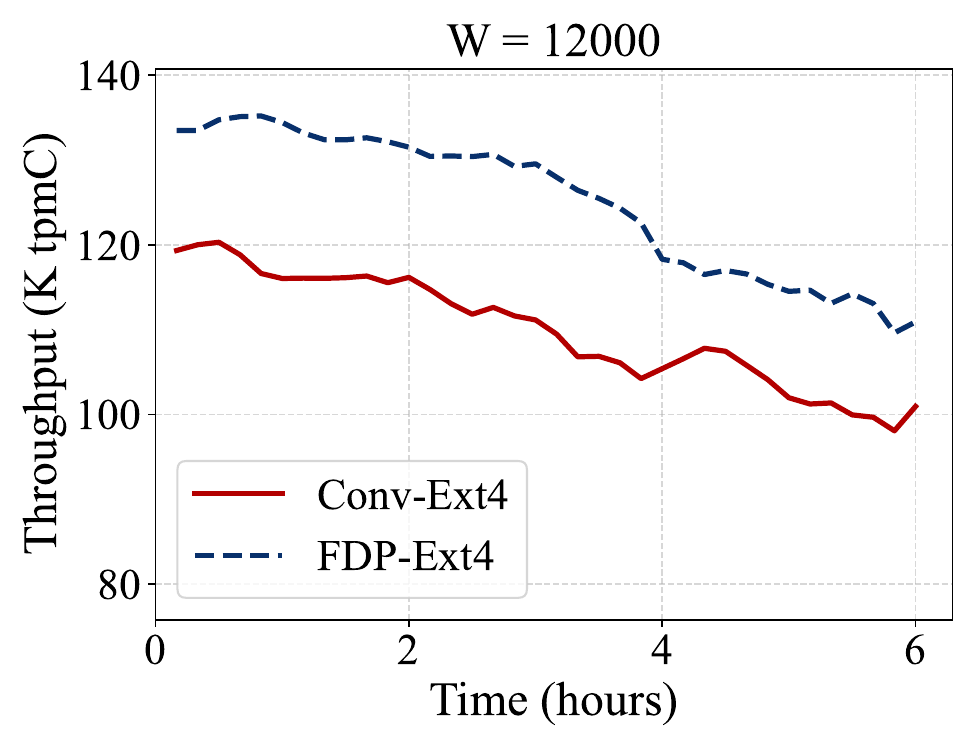}
    \end{minipage}
\caption{TPC-C throughput over time for Conv and FDP with 6000 (left) and 12000 (right) warehouses.}

    \label{fig:throughput_warehouse}
\end{figure}

% \bluetext{ Figure~\ref{fig:waf_warehouse} shows WAF as device utilization increases at $W=6000$ and $W=12000$. At $W=6000$, both configurations maintain a WAF of approximately 1.0 through 30\% utilization. At 40\%, Conv's WAF rises to 1.19, while FDP's remains at 1.00. The relative WAF reduction achieved by FDP peaks at approximately 34\% at 80\% utilization (1.23 versus 1.86), then declines to 13\% at 90\% (1.77 versus 2.04). At $W=12000$, Conv's WAF begins to rise at 50\% utilization, while FDP's remains approximately 1.0 through 70\%. The relative WAF reduction peaks at approximately 36\% at 70\% utilization (1.01 versus 1.59), then declines to 10\% at 90\% (1.94 versus 2.16). The reduction thus peaks at a lower utilization than at $W=6000$. However, comparisons across database sizes should account for differences in workload phase, as the LOAD phase ends at approximately 36\% utilization for $W=6000$ and 70\% for $W=12000$. Consequently, measurements at the same utilization may represent transaction processing in one experiment and data loading in the other.  Both database sizes exhibit the same non-monotonic trend, with FDP's relative WAF reduction increasing initially and diminishing near saturation. FDP nevertheless maintains a lower WAF than Conv at 90\% utilization. The lower WAF is accompanied by higher throughput (Figure~\ref{fig:throughput_warehouse}). FDP achieves 9.7\% higher average TpmC at $W=6000$ (158,757 versus 144,736) and 13.7\% higher average TpmC at $W=12000$ (124,998 versus 109,895). }

 Figure~\ref{fig:waf_warehouse} shows WAF as device utilization increases at $W=6000$ and $W=12000$. At $W=6000$, both configurations maintain a WAF of approximately 1.0 through 30\% utilization. At 40\%, Conv's WAF rises to 1.19, while FDP's remains at 1.00. FDP's relative WAF reduction peaks at approximately 34\% at 80\% utilization (1.23 versus 1.86), then declines to 13\% at 90\% (1.77 versus 2.04). At $W=12000$, Conv's WAF begins to rise at 50\% utilization, while FDP's remains approximately 1.0 through 70\%. The relative reduction peaks at approximately 36\% at 70\% utilization (1.01 versus 1.59), then declines to 10\% at 90\% (1.94 versus 2.16). Comparisons across database sizes should account for the different LOAD-phase boundaries (approximately 36\% utilization at $W=6000$ and 70\% at $W=12000$), as the same utilization may correspond to different workload phases. Both database sizes exhibit a non-monotonic trend, with FDP's relative WAF reduction increasing initially and diminishing near saturation. Figure~\ref{fig:throughput_warehouse} shows that these WAF reductions are accompanied by higher throughput. FDP improves average TpmC by 9.7\% at $W=6000$ (158\,757 versus 144\,736) and 13.7\% at $W=12000$ (124\,998 versus 109\,895).

\subsubsection{Filesystem Sensitivity on MySQL}
\label{sec:mysql_fs_sensitivity}

 Figure~\ref{fig:mysql_fs_waf} shows that ext4 and XFS exhibit similar WAF under Conv throughout the run (median relative difference $\approx$1\%, with occasional spikes up to $\approx$3.6\%), ending at 1.82 and 1.86, respectively. Under FDP, the two curves remain close for approximately the first 6 hours ($\sim$24\,TiB of cumulative host writes from the start of the LOAD phase, equivalent to $\sim$7$\times$ the device capacity), after which the gap widens. At approximately 7.7 hours ($\sim$32\,TiB of cumulative host writes, 62--64\% device utilization), XFS's WAF is 6.1\% higher than ext4's. It subsequently increases more rapidly, while ext4's WAF rises more gradually. By the end of the run, FDP's interval WAF reaches 1.28 on ext4 and 1.59 on XFS, with ext4's WAF 19.3\% lower than XFS's. Relative to the corresponding Conv baselines, these values represent WAF reductions of 29.6\% on ext4 and 14.9\% on XFS. Thus, FDP provides a larger WAF reduction on ext4 in this experiment, with the advantage emerging primarily during the later, higher-utilization portion of the run. We therefore use ext4 as the baseline filesystem for the MySQL experiments.

\subsubsection{Discard is Not a Confounding Variable}
\label{sec:mysql_discard_control}

\begin{figure}[t]
  \centering
  \begin{minipage}[t]{0.42\linewidth}
    \centering
    \includegraphics[width=\linewidth]{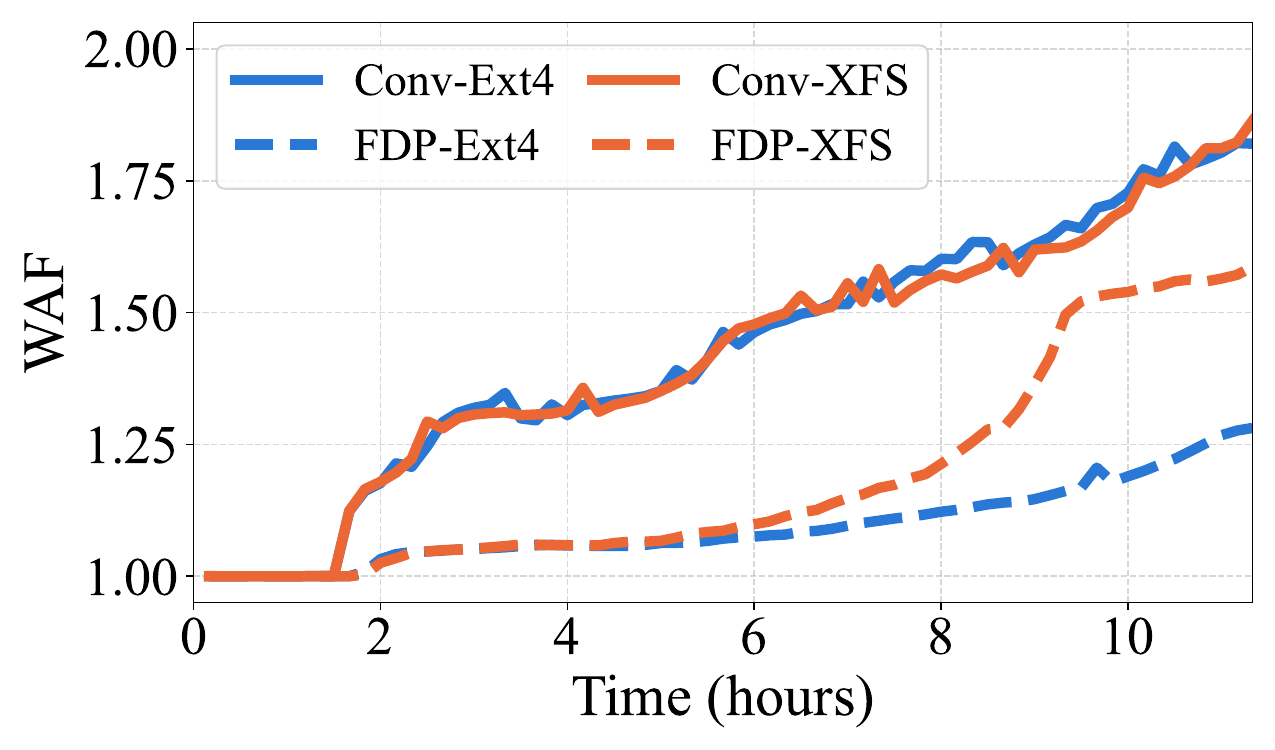}
    \captionof{figure}{WAF on ext4 and xfs.}
    \label{fig:mysql_fs_waf}
  \end{minipage}
  \hfill
  \begin{minipage}[t]{0.42\linewidth}
    \centering
    \includegraphics[width=\linewidth]{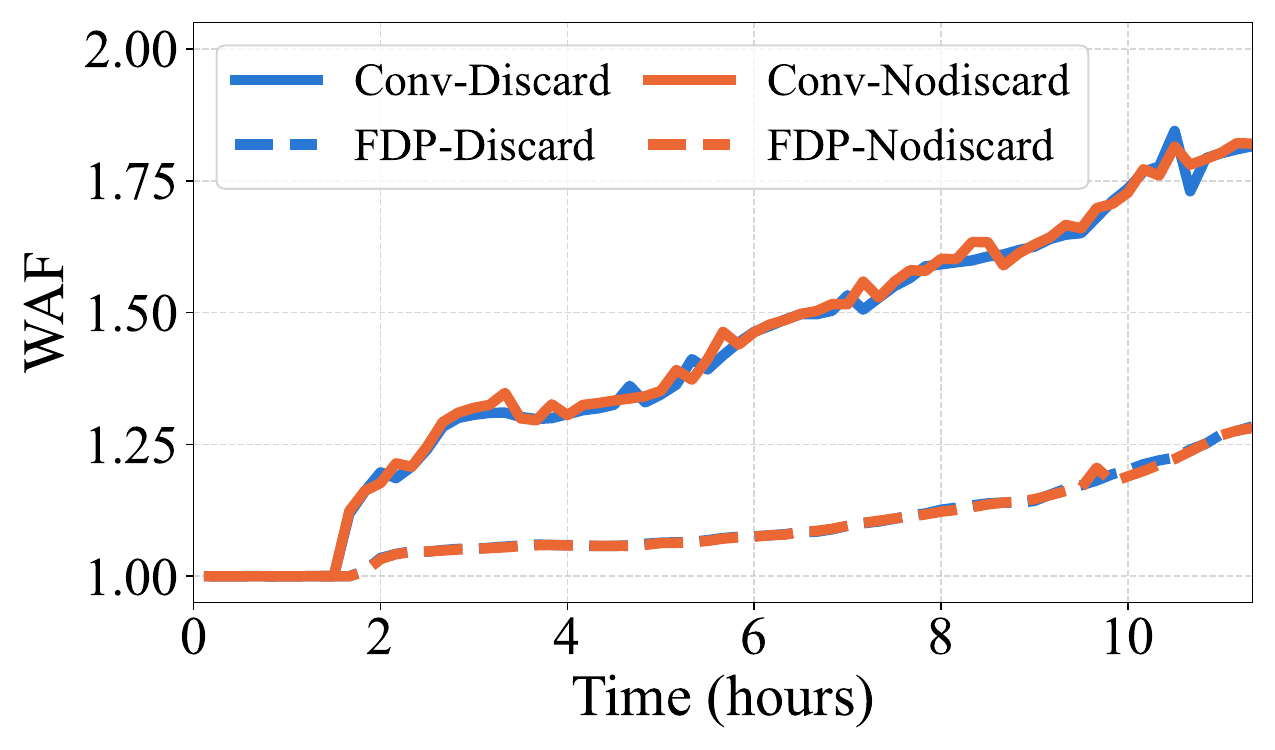}
    \captionof{figure}{ WAF with vs.\ without discard.}
    \label{fig:mysql_discard_control}
  \end{minipage}
\end{figure}

 Figure~\ref{fig:mysql_discard_control} evaluates the effect of the \texttt{discard} mount option by comparing ext4 with and without discard under both Conv and FDP. The two Conv curves closely track each other, with a median relative difference of 0.5\% and a maximum transient difference of 3.4\%. The FDP curves are even closer, with a median relative difference of 0.1\%. For each placement mode, the final WAF values differ by less than 2\% between discard settings, reaching approximately 1.82 for Conv and 1.28 for FDP. FDP reduces WAF relative to Conv by 29.3\% with discard and 29.6\% without it. These results indicate that the discard mount option has little effect on WAF under this workload and does not materially affect FDP's benefit on ext4. 

\subsubsection{Multi-Tenant MySQL Deployment}
In multi-tenant scenarios, each MySQL instance adopts the placement Policy~3 described in Table~\ref{tab:mysql_policy}. Files in the same placement class share an RUH across instances, preserving separation between file classes without requiring additional RUHs for each tenant.
We evaluate two and four tenants with 3000 and 1500 warehouses per instance, respectively, both totaling 6000 warehouses and reaching $\sim$70\% SSD utilization representative of production workloads.

\begin{figure}[t]
    \centering
    \begin{subfigure}{0.48\linewidth}
        \centering
        \includegraphics[width=\linewidth]{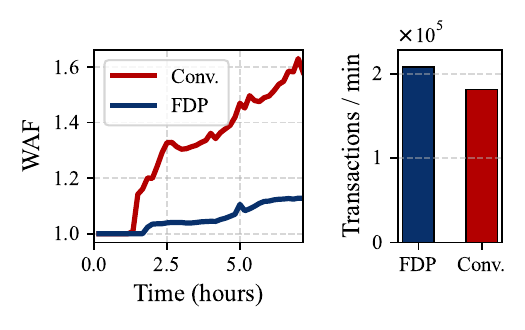}
        \caption{Two-instance workload}
        \label{fig:mysql_dual}
    \end{subfigure}
    \hfill
    \begin{subfigure}{0.48\linewidth}
        \centering
        \includegraphics[width=\linewidth]{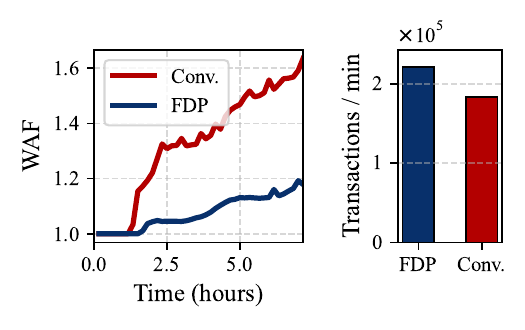}
        \caption{Four-instance workload}
        \label{fig:mysql_quad}
    \end{subfigure}
\caption{WAF under multi-tenant MySQL workloads.
         FDP achieves 12--16\% WAF reduction across both configurations.}
    \label{fig:mysql_multi}
   %\vspace{-4ex}
\end{figure}

Figure~\ref{fig:mysql_multi} shows that FDP reduces WAF from 1.58 to 1.13 with two tenants (28.7\%) and from 1.64 to 1.18 with four tenants (27.9\%) by the end of the run. The similar reductions indicate that FDP retains its benefit as the tenant count increases from two to four at a fixed aggregate database size. Conv's WAF exceeds 1.10 at approximately 1.5 hours in both configurations, whereas FDP crosses the same threshold at approximately 5 hours with two tenants and 4.3 hours with four tenants. This delay of approximately 2.8--3.5 hours suggests that separating file classes helps limit WAF growth as device utilization rises. These WAF reductions are accompanied by higher average host write bandwidth during the 6-hour RUN phase. With two tenants, bandwidth increases from 1517 MB/s under Conv to 1709 MB/s under FDP (12.7\%). With four tenants, it increases from 1531 MB/s to 1837 MB/s (20.0\%).

% FDP sustains higher host write bandwidth than Conv during the 6-hour RUN phase. With two tenants, average bandwidth increases from 1517 MB/s under Conv to 1709 MB/s under FDP (12.7\%). With four tenants, it increases from 1531 MB/s to 1837 MB/s (20.0\%). After the initial transient period, Conv's bandwidth declines by 8.2\% and 10.4\% with two and four tenants, respectively, compared with 1.2\% and 4.0\% under FDP. These results are consistent with FDP's lower WAF and reduced GC overhead, leaving more device bandwidth available for host writes.

The MySQL experiments demonstrate that FDP reduces device-level WAF with only a few placement streams for InnoDB's mix of B-tree page updates and log writes. Together with the RocksDB results, these findings support the applicability of lifetime-based placement across LSM-tree and B-tree storage architectures.

%\subsection{Multi-tenant Deployments}

\section{Conclusion}
NVMe FDP is an industry-driven shift in the SSD interface landscape whose backward compatibility and ability to reduce WAF with non-invasive software changes is driving adoption in datacenter deployments. With stable Linux support for I/O Passthru and emerging filesystem-based lifetime hints, FDP enables practical data separation in widely deployed storage systems. We show that RocksDB’s out-of-place architecture can be leveraged for WAF gains to separate SST and WAL files with different lifetimes using both filesystem and I/O Passthru interfaces.  Similar gains can also be obtained in MySQL's predominantly in-place update architecture by separating data from logs, table data, and metadata using filesystem hints. Across synthetic and real-world workloads, we show that FDP delivers near-ideal WAF in RocksDB and achieves a 10–15\% WAF reduction in MySQL at high SSD utilization under both single-tenant and multi-tenant settings. %Our work highlights FDP’s potential for predictable performance and improved storage efficiency in modern datacenters.

%\clearpage
\newpage
\bibliographystyle{ACM-Reference-Format}
\bibliography{fdp}
\conditionaltext{}{\appendix
\newpage
\section{The Promise of the Raw FDP Device (Contd.)}
\label{appendix:microbench}

\subsection{Skew vs WAF}
\label{app:skew-vs-waf}

Figure~\ref{fig:skew_baseline} shows WAF as a function of the total amount of data written by the user (TiB) for a single workload of Zipfian random writes at eight skew levels ($skew \in \{0, 0.2, 0.4, 0.6, 0.8, 1.0, 1.2, 1.4\}$).

At low skew values ($skew \leq 0.6$), writes are distributed nearly uniformly across the LBA space, causing WAF to rise steeply and saturate at approximately 4.2 after 15--20\,TiB of user writes (approximately 4--5$\times$ the device capacity). This is consistent with the behaviour of a nearly full device under uniform random writes. As skew increases ($skew \in \{0.8, 1.0\}$), writes concentrate on a smaller hot region. The resulting locality reduces the fraction of cold data intermixed with hot writes during GC, and WAF saturates at progressively lower levels of approximately 1.6 and 1.2, respectively. At high skew ($skew \geq 1.2$), writes are highly concentrated, and WAF remains close to 1.0 throughout the entire write range, indicating that GC operates almost exclusively on invalidated hot pages with minimal migration of valid pages.

These results establish the relationship between access skew and device WAF under a single workload. When natural locality is insufficient (i.e., at low skew), FDP can enforce physical separation between hot and cold data streams to achieve comparable WAF reductions, as evaluated in the mixed workload experiments in \S~\ref{sec:workload-skew}.

\label{app:skew-vs-waf}
\begin{figure}[htbp]
    \centering
    \includegraphics[width=0.9\linewidth]{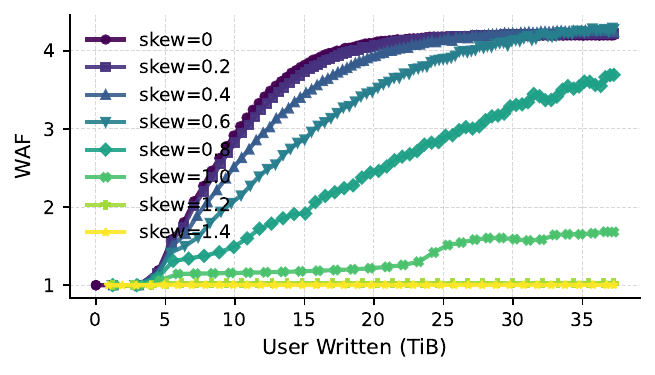}
    \caption{Baseline WAF evolution over time (User Written TiB) for a single workload under varying Zipfian skewness parameters. Higher skewness (e.g., $skew=1.4$) maintains near-ideal WAF, while uniform random workloads ($skew=0$) result in high write amplification.}
    \label{fig:skew_baseline}
\end{figure}

\subsection{Homogeneous vs. Heterogeneous RUH Placement}
\label{app:homo_heter_map}

\textbf{Setup.}
This experiment builds directly on \S~\ref{sec:seq-workloads} and retains the same workload construction and logical layout. Each logical region is subdivided into workloads of equal size that run in parallel, preserving the aggregate write volume and access characteristics.
We evaluate four concurrency levels (24, 32, 48, and 64 jobs) to examine how sensitivity to RUH placement changes with increasing parallelism.

We compare three configurations.
\begin{itemize}
    \item \textbf{Conv} uses a conventional namespace without FDP support and serves as the baseline.
    \item \textbf{U-FDP (Homogeneous Placement)} enables FDP with uniform placement, assigning each RUH write streams with equally sized LBA ranges.
    \item \textbf{H-FDP (Heterogeneous Placement)} enables FDP with heterogeneous placement, assigning each RUH write streams with maximally diverse LBA range sizes. The ratio of the largest to the smallest range within a RUH reaches 11.5$\times$.
\end{itemize}
All three configurations use identical logical offsets, total write volumes, and access patterns. U-FDP and H-FDP use the same LBA ranges for each job. They differ only in the \texttt{fdp\_pli} assignment, which determines the RUH to which each write stream is directed.

\begin{figure*}[htbp]
    \centering
    \begin{subfigure}{0.24\textwidth}
        \centering
        \includegraphics[width=\linewidth]{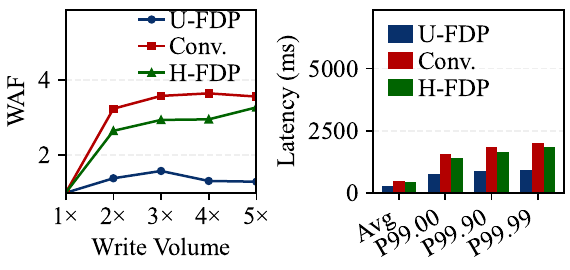}
        \caption{WL24}
        \label{fig:wl24}
    \end{subfigure}
    \hfill
    \begin{subfigure}{0.24\textwidth}
        \centering
        \includegraphics[width=\linewidth]{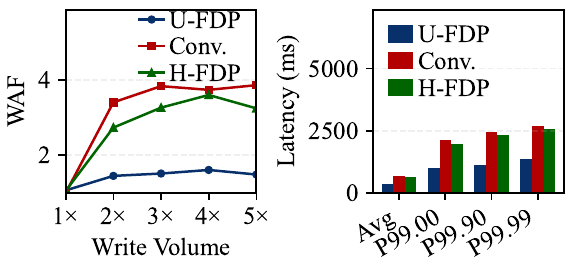}
        \caption{WL32}
        \label{fig:wl32}
    \end{subfigure}
    \hfill
    \begin{subfigure}{0.24\textwidth}
        \centering
        \includegraphics[width=\linewidth]{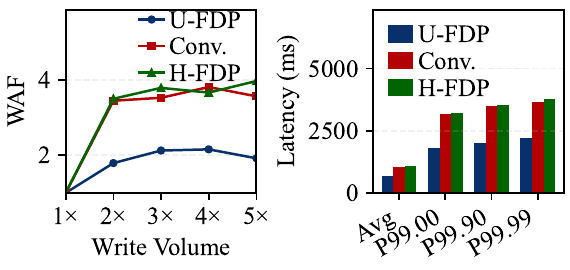}
        \caption{WL48}
        \label{fig:wl48}
    \end{subfigure}
    \hfill
    \begin{subfigure}{0.24\textwidth}
        \centering
        \includegraphics[width=\linewidth]{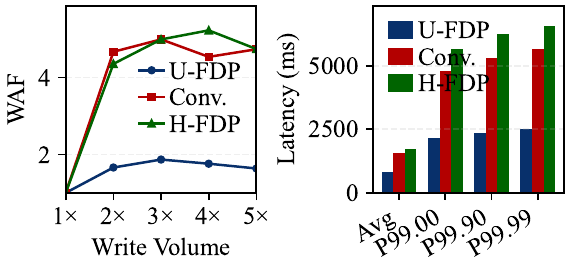}
        \caption{WL64}
        \label{fig:wl64}
    \end{subfigure}
\caption{Impact of RUH placement under increasing concurrency.
The left panels show write amplification versus cumulative write volume normalized to device capacity.
The right panels show average and tail latencies through P99.99.
Homogeneous placement (U-FDP) consistently maintains lower amplification and stable latency.
As concurrency increases, heterogeneous placement produces behavior progressively closer to that of conventional SSDs.}
\label{fig:ruh_scaling}
\end{figure*}

\textbf{Results.}
Figure~\ref{fig:ruh_scaling} summarizes the impact of RUH placement under increasing concurrency (24--64 jobs).

Across all levels of parallelism, U-FDP consistently maintains lower write amplification and more stable tail latency than both H-FDP and the conventional baseline (Conv). However, sensitivity to RUH assignment increases markedly with concurrency.

At moderate concurrency (WL24--WL32), H-FDP still provides benefits over Conv, although differences in LBA range sizes within each RUH already increase write amplification. By WL48, this advantage effectively disappears. H-FDP exhibits write amplification and tail latency nearly indistinguishable from the baseline, indicating that improper stream colocation largely negates FDP's isolation benefits at this scale.

The trend persists at WL64, confirming that under extreme parallelism, RUH mapping quality becomes the dominant factor determining whether FDP isolation remains effective. In contrast, homogeneous placement continues to limit WAF growth and maintain predictable latency across all concurrency levels.

\textbf{Discussion.}
Two consistent trends emerge across the concurrency levels. First, RUH placement quality directly affects write amplification. Homogeneous placement systematically reduces imbalance within each RUH and preserves FDP's isolation benefits. Second, sensitivity to placement increases with parallelism. The performance gap between homogeneous and heterogeneous placement widens from WL24 to WL64. Under extreme concurrency, improper stream colocation largely nullifies FDP's advantage.

\takeaway{
FDP provides strong isolation benefits, but its effectiveness increasingly depends on RUH assignment quality as concurrency grows. Grouping streams with equally sized LBA ranges within a RUH preserves scalability, whereas heterogeneous colocation produces behavior increasingly similar to that of conventional SSDs under high parallelism.
}

\subsection{Scalability of FDP Across Device Sizes}
\label{sec:app-fdp-device-sizes}
\textbf{Experimental setup.}
We repeat the uneven sequential fio benchmark from Section~\ref{sec:seq-workloads} on three NVMe devices with capacities of 1.88\,TB, 3.84\,TB, and 7.52\,TB, varying the number of jobs per RUH across 2, 4, 6, and 8.

\textbf{Results.}
Figure~\ref{fig:waf_device} reports WAF at steady state for U-FDP and Conv across all three devices.

On the 3.84\,TB device, Conv WAF rises from 2.99 to 3.97 as concurrency increases, a 33\% increase driven by interference between streams. U-FDP mitigates this pressure through explicit placement, reducing WAF by 49\% at peak load. On the 1.88\,TB device, Conv WAF remains largely stable across all concurrency levels (2.99--3.00), while U-FDP consistently reduces WAF by 40\%. This indicates that placement isolation remains beneficial even when concurrency does not substantially increase baseline WAF.

The 7.52\,TB device exhibits the lowest absolute WAF under both configurations. Conv WAF rises modestly from 2.43 to 2.79, while U-FDP maintains WAF at approximately 1.0--1.13 throughout, yielding a 59\% reduction at peak load. The larger address space limits the magnitude of interference but does not eliminate the advantage of explicit placement control.

\begin{figure}[t]
    \centering
    \includegraphics[width=0.9\linewidth]{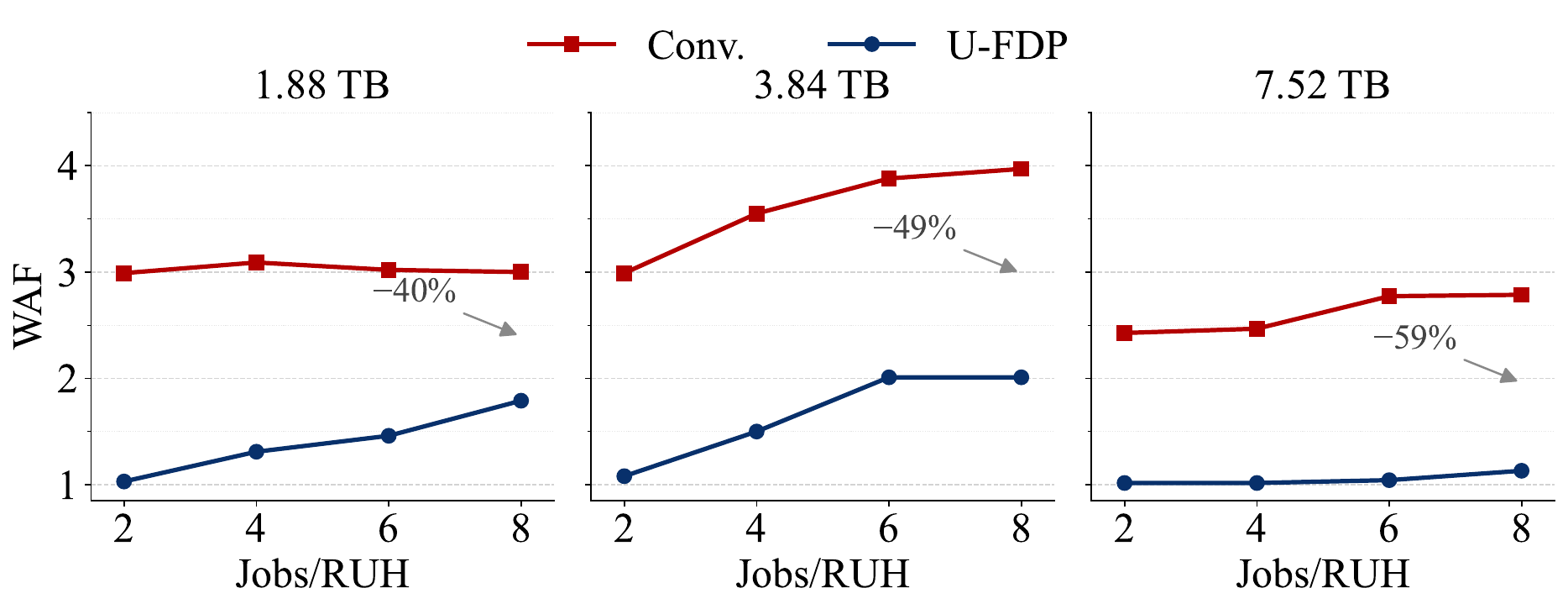}
    \caption{Steady-state WAF under U-FDP and Conv as a function of
    per-RUH job count across three devices of increasing capacity.
    Annotations indicate WAF reduction at maximum concurrency
    (8\,jobs/RUH).}
    \label{fig:waf_device}
\end{figure}

\takeaway{U-FDP consistently reduces WAF across all device sizes and concurrency levels, with reductions of 40\% to 59\% at peak load. Larger devices limit the magnitude of interference through their larger physical address space, but explicit placement control remains effective across capacities. These results suggest that FDP is broadly applicable, including in configurations without capacity constraints.}

\textbf{Contract.}
The physical address space available to each stream governs the magnitude of interference but does not replace explicit placement control. U-FDP consistently reduces WAF across the full range of capacities tested.

%\textbf{FDP contract.}
%Under high concurrency, treat RUH assignment as a first-class placement policy. 
%When workloads must share a RUH, colocate streams with similar sizes or write intensities to preserve FDP’s isolation advantage.
% \subsection{{Impact of FDP on WAF with Hot/Cold Data Separation}}

\subsection{Multiple Containers with Similar Filesystem Workloads}

% \usepackage{graphicx}
\begin{table}[htbp]
\resizebox{\columnwidth}{!}{%
\begin{tabular}{ccccl}
\hline
Write Type & Block Size (KB) & File Size (MB) & Write Hint & \# Files \\ \hline
Sequential & 128             & 100            & Short      & 1200     \\ \hline
Sequential & 64              & 64             & Medium     & 1800     \\ \hline
Random     & 8               & 24             & Long       & 4000     \\ \hline
Random     & 4               & 24             & extreme    & 2000     \\ \hline
\end{tabular}%
}
\caption{FIO Write Type Configurations for FileSystem}
\label{tab:fio-fs}
\end{table}

\textbf{Setup.}
We evaluate how FDP interacts with filesystem allocation policies under workloads with mixed data lifetimes. We modify the Linux kernel to propagate \texttt{write\_hint} values associated with files or inodes to Placement Identifiers (PLIs) at the device level, enabling placement based on data lifetime without altering filesystem internals.

All experiments use FIO with \texttt{io\_uring} and direct I/O. To eliminate layout bias, we first precondition the SSD with large sequential writes to regions of 2.0~TB and 1.3~TB, followed by three hours of random overwrites with mixed write sizes to establish fragmentation at steady state. We then run a workload for eight hours with four concurrent lifetime classes, namely \emph{short} (128~KB sequential), \emph{medium} (64~KB sequential), \emph{long} (8~KB sequential), and \emph{extreme} (4~KB random). Each class writes to independent directories and is tagged with a distinct \texttt{write\_hint}, allowing FDP to separate data by lifetime during placement. We run the same workload on EXT4, XFS, Btrfs, and F2FS to isolate the effects of each filesystem.

\textbf{Results.}
\emph{Write amplification.} Figure~\ref{fig:waf_fs_comparison} shows that FDP consistently reduces WAF across all filesystems. Under the conventional SSD configuration (Conv), WAF increases to approximately 1.8 as the device fills, reflecting interference between data with different lifetimes within erase blocks. In contrast, FDP configurations stabilize near 1.0, indicating that physical write volume nearly matches logical write volume even with heterogeneous update granularities.

\label{sec:app:fs-microbench}
\begin{figure}[htbp]
    \centering
    \begin{subfigure}[b]{0.48\linewidth}
        \centering
        \includegraphics[width=\linewidth]{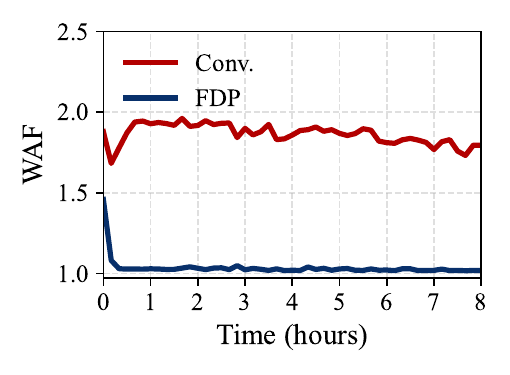}
        \caption{EXT4}
    \end{subfigure}
    \hfill
    \begin{subfigure}[b]{0.48\linewidth}
        \centering
        \includegraphics[width=\linewidth]{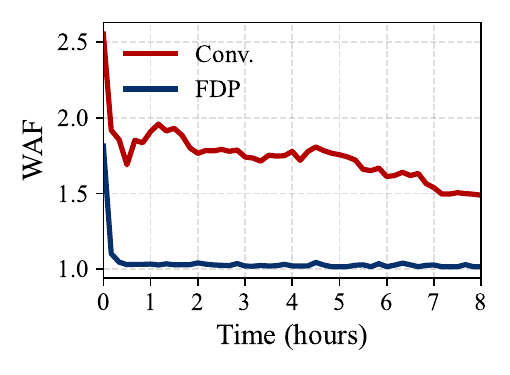}
        \caption{XFS}
    \end{subfigure}
    % \vskip 1em
    \begin{subfigure}[b]{0.48\linewidth}
        \centering
        \includegraphics[width=\linewidth]{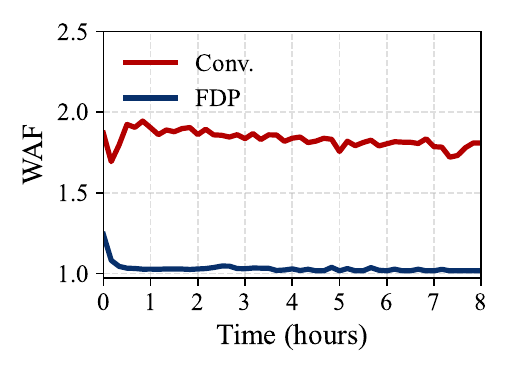}
        \caption{Btrfs}
    \end{subfigure}
    \hfill
    \begin{subfigure}[b]{0.48\linewidth}
        \centering
        \includegraphics[width=\linewidth]{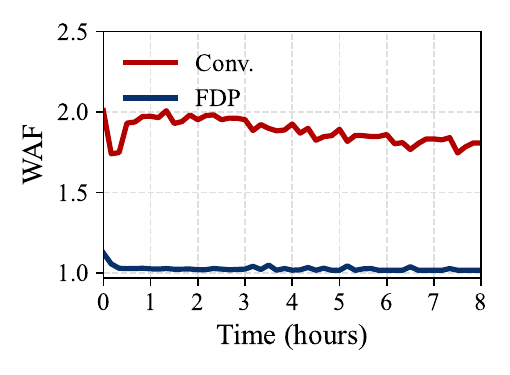}
        \caption{F2FS}
    \end{subfigure}
    \caption{Comparison of WAF across four Linux filesystems with and without FDP. FDP consistently reduces write amplification across all filesystems.}
    \label{fig:waf_fs_comparison}
\end{figure}

\begin{figure}[htbp]
    \centering
    \begin{subfigure}[b]{0.49\linewidth}
        \includegraphics[width=\linewidth]{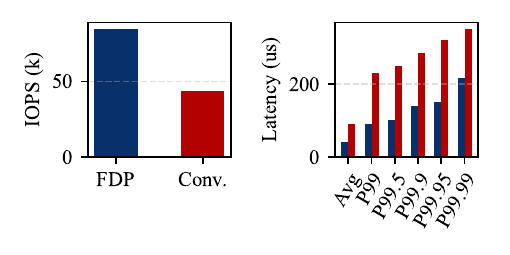}
        \caption{EXT4}
    \end{subfigure}
    \hfill
    \begin{subfigure}[b]{0.49\linewidth}
        \includegraphics[width=\linewidth]{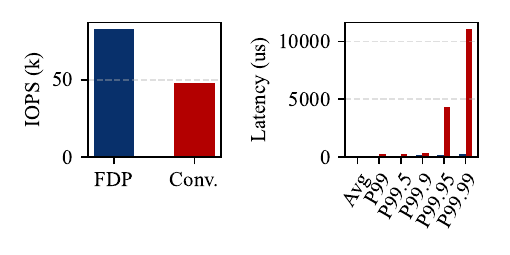}
        \caption{XFS}
    \end{subfigure}
    \hfill
    \begin{subfigure}[b]{0.49\linewidth}
        \includegraphics[width=\linewidth]{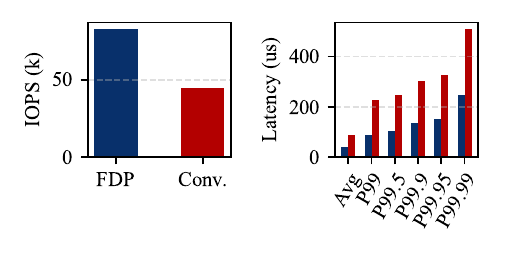}
        \caption{Btrfs}
    \end{subfigure}
    \hfill
    \begin{subfigure}[b]{0.49\linewidth}
        \includegraphics[width=\linewidth]{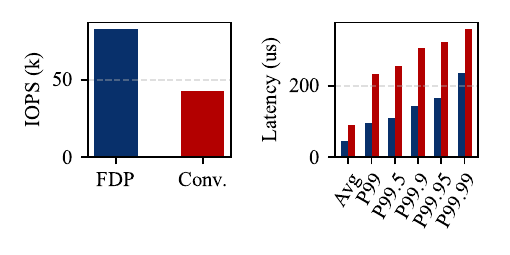}
        \caption{F2FS}
    \end{subfigure}
    \caption{Comparison of latency and IOPS across filesystems.}
    \label{fig:fs_metric_comparison}
\end{figure}

\emph{Performance.} Figure~\ref{fig:fs_metric_comparison} presents latency and throughput results. EXT4, Btrfs, and F2FS show moderate improvements under FDP, with P99.99 latency reduced by 34--51\% while maintaining similar IOPS. XFS shows substantially larger gains, achieving approximately 83~K IOPS compared with 47~K under Conv, along with consistently lower tail latency across percentiles. We attribute this improvement to the allocation group architecture of XFS, which aligns naturally with FDP's placement model based on streams and better preserves spatial isolation across data classes.

\takeaway{Across diverse filesystems, FDP consistently restores efficient placement based on data lifetime under mixed workloads. Conventional SSDs suffer from interference between data with different lifetimes, increasing WAF and tail latency. FDP isolates these heterogeneous streams during placement, stabilizing WAF near 1.0. The gains are particularly pronounced when filesystem allocation structures, such as XFS allocation groups, align with FDP's placement granularity. This highlights the importance of compatibility between filesystem layout and stream isolation at the device level.}

\section{RocksDB Evaluation (Contd.)}

\subsection{Update-Intensive Workloads with Different read-update Ratio}
\label{app:update_qos_2080}
\begin{figure}[htbp]
    \centering
    \begin{subfigure}{0.48\linewidth}
        \centering
        \includegraphics[width=\linewidth]{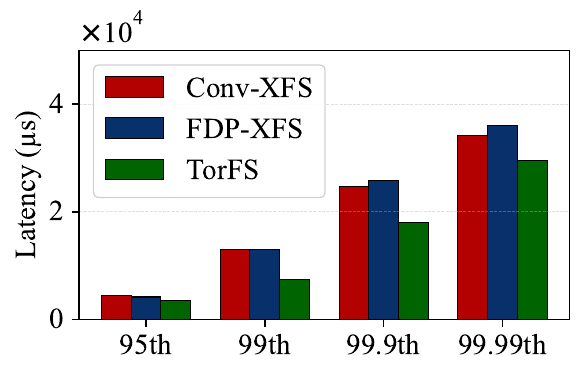}
        \caption{Read latency}
        \label{fig:qos_read_2080}
    \end{subfigure}
    \hfill
    \begin{subfigure}{0.48\linewidth}
        \centering
        \includegraphics[width=\linewidth]{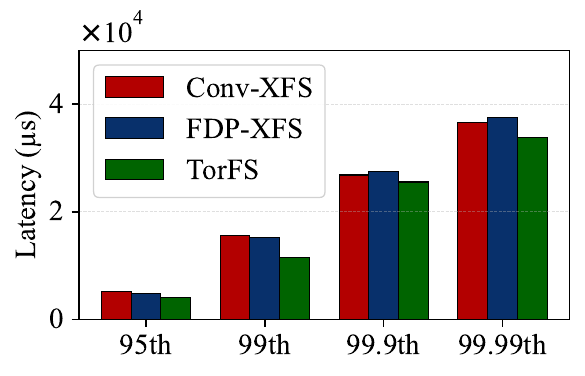}
        \caption{Update latency}
        \label{fig:qos_update_2080}
    \end{subfigure}
    \caption{Tail latency percentiles for 20\% read / 80\% update workload. Under high update pressure, FDP maintains stable tail latencies comparable to baseline performance.}
    \label{fig:qos_2080}
\end{figure}

Figures~\ref{fig:qos_read_2080} and~\ref{fig:qos_update_2080} show the tail latency distributions for read and update operations, respectively, under the 20/80 workload dominated by updates. Consistent with the 50/50 results in \S~\ref{sec:evaluation-update}, all three configurations (Conv-XFS, FDP-XFS, and TorFS) exhibit comparable latency at all measured percentiles (95th, 99th, 99.9th, and 99.99th) for both operation types. FDP reduces WAF from 1.45 to 1.05 under this workload without introducing an observable latency penalty. These results confirm that FDP's placement isolation mechanism does not introduce queuing delays or scheduling artifacts even as update intensity increases, supporting its applicability to workloads sensitive to latency.

\subsection{Tail Latency Under the SSD Utilization Stress Test}
\label{app:ssd_uti_stress_latency}

Figure~\ref{fig:qos_large} shows the tail latency distributions for READ and UPDATE operations under YCSB Workload~A at high device utilization. All three approaches exhibit broadly comparable latency at p95, p99, and p99.9 for both operation types.

At p99.99, FDP-XFS and TorFS reduce READ latency relative to Conv-XFS by 26.9\% and 25.9\%, respectively (50\,591~µs and 51\,263~µs versus 69\,183~µs). They also reduce UPDATE latency by 20.7\% and 18.4\%, respectively (56,223~µs and 57\,887~µs versus 70\,911~µs). These improvements at the extreme tail stem from FDP's cleaner data placement within each RU, which reduces the frequency and severity of I/O stalls caused by GC. Such stalls dominate the worst observed latencies at high device utilization. For deployments sensitive to latency where meeting p99.99 SLA requirements is critical, FDP provides a meaningful advantage even as available SSD space nears exhaustion.

\begin{figure}[t]
    \centering
    \begin{subfigure}{0.48\linewidth}
        \centering
        \includegraphics[width=\linewidth]{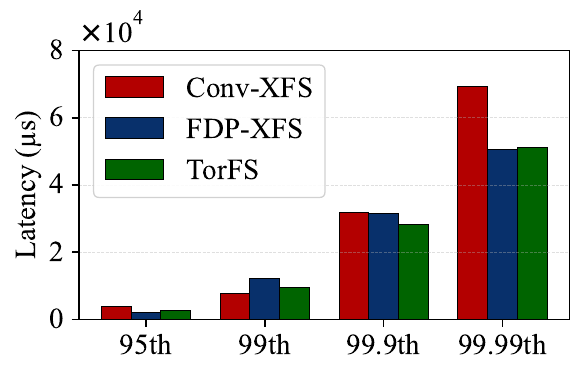}
        \caption{Read latency}
        \label{fig:qos_read_large}
    \end{subfigure}
    \hfill
    \begin{subfigure}{0.48\linewidth}
        \centering
        \includegraphics[width=\linewidth]{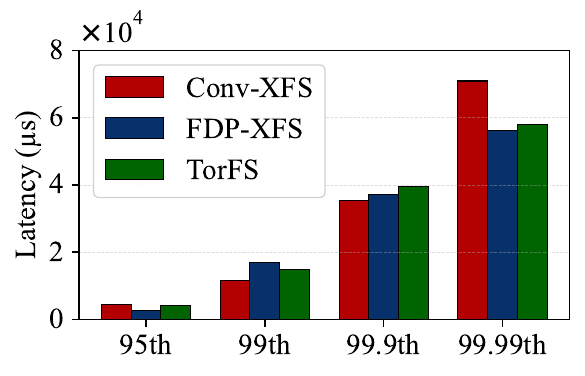}
        \caption{Update latency}
        \label{fig:qos_update_large}
    \end{subfigure}
        \caption{Tail latency under large-scale stress test.}
       %\vspace{-6pt}
    \label{fig:qos_large}
\end{figure}

\section{TorFS vs UFROP}
\label{app:TorFS_vs_ufrop}

{Traditional file systems adopt a layered architecture in which data traverses the VFS, the file system, and the block layer before reaching the device. This separation provides generality and portability but introduces software stack overhead that can become a bottleneck when performance is critical. Storage systems designed around direct application access instead bypass the kernel file system entirely, allowing applications to write using native device semantics and eliminating this redundant processing.}

Recent work applies this approach to exploit FDP semantics in RocksDB. UFROP~\cite{ufrop} and TorFS are both RocksDB plugins for FDP SSDs that share two design principles. They use hints from RocksDB to separate data into multiple streams and bypass the kernel file system to issue direct I/O. They differ in how they lay out files on the device. UFROP partitions the user data region into blocks of a fixed size and maps each file through direct and indirect blocks. A file that fits within one block is stored contiguously, while larger files use indirect blocks. Unused space within a block cannot be shared with another file, resulting in internal fragmentation. The bounded number of indirect blocks per file also limits the maximum file size. TorFS instead allocates segments of variable length to match the actual file size, eliminating internal fragmentation without imposing an intrinsic limit on file size.

\begin{figure}[t]
  \centering
  \includegraphics[width=0.8\columnwidth]{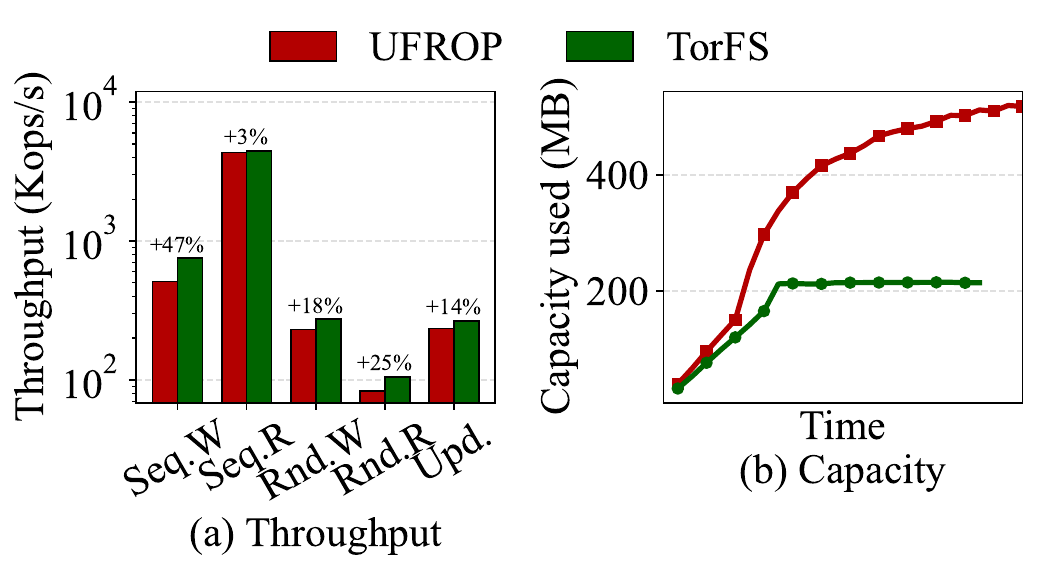}
 % \vspace{-4mm}
  \caption{{TorFS versus UFROP on \texttt{db\_bench}: (a) throughput per workload and (b) on-storage space consumed over time. Percentages are relative to UFROP.}}
  \vspace{-3mm}
  \label{fig:TorFS_vs_ufrop_thr_cap}
\end{figure}

\begin{figure}[t]
  \centering
  \includegraphics[width=0.8\columnwidth]{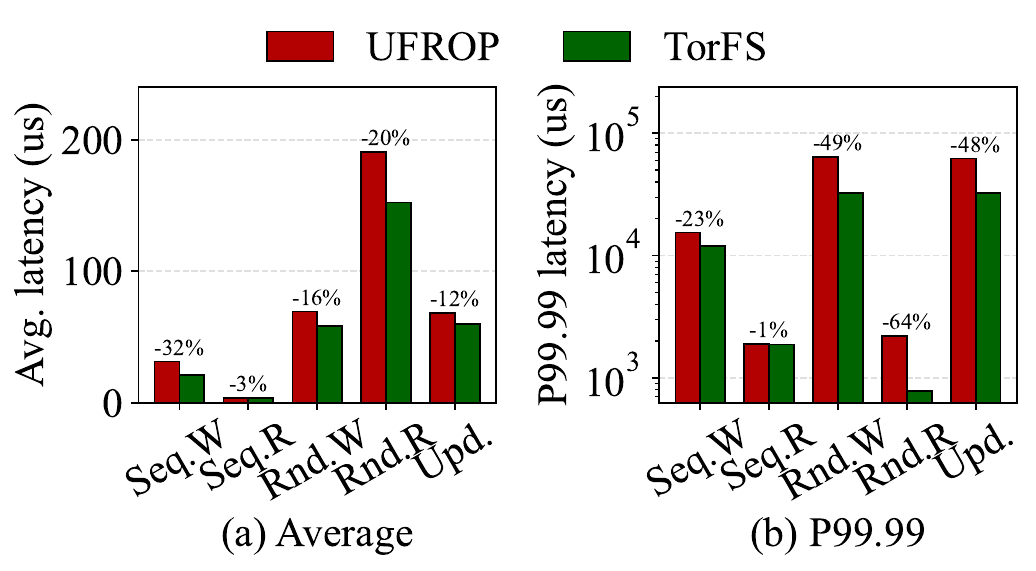}
 % \vspace{-4mm}
  \caption{{Latency of TorFS and UFROP on \texttt{db\_bench}: (a) average and (b) P99.99. Percentages are relative to UFROP.}}
 % \vspace{-3mm}
  \label{fig:TorFS_vs_ufrop_lat}
\end{figure}

\textbf{Performance.} We compare the two plugins using \texttt{db\_bench} with 20 million records per thread across 16 threads under sequential read, sequential write, random read, random write, and update workloads. Figure~\ref{fig:TorFS_vs_ufrop_thr_cap}(a) shows nearly identical sequential read performance, with TorFS achieving a 3\% improvement, since neither plugin lies on the critical path for reads. TorFS consistently outperforms UFROP under workloads dominated by writes. Sequential write throughput increases by 47\%, while average and P99 latency decrease by 32\% and 34\%, respectively. Update throughput increases by 14\%, with a 52\% reduction in P99 latency (Figure~\ref{fig:TorFS_vs_ufrop_lat}). Random write and read throughput improve by 18\% and 25\%, respectively, while extreme tail latency falls by roughly half for random writes and updates. These gains stem from TorFS's I/O submission strategy. UFROP submits requests individually, whereas TorFS batches them, reducing the number of submissions to the device and the associated overhead per request.

\textbf{Space efficiency.} Under the same workload, UFROP consumes 518~MB of physical space, compared with only 214~MB for TorFS, a factor of $2.4\times$ (Figure~\ref{fig:TorFS_vs_ufrop_thr_cap}(b)). TorFS's space consumption stabilizes once the LSM tree reaches steady state, whereas UFROP's continues to grow as fragmented blocks accumulate. This difference directly reflects the elimination of internal fragmentation through allocation based on segments. Together, these performance and space efficiency results motivate our choice of TorFS as the storage plugin.

}

\end{document}